\documentclass[12pt]{article}
\pdfoutput=1
\usepackage[utf8]{inputenc}
\usepackage[left=2.55cm, right=2.55cm, top=2.55cm, bottom=2.55cm]{geometry}
\usepackage{tikz}
\usepackage{tikz-feynman}
\usepackage{contour}
\usepackage{amsmath,amssymb,amsbsy}
\usepackage{slashed}
\usepackage{xcolor}
\usepackage{graphicx}
\usepackage{tikz}
\usepackage{tikz-feynman}
\usepackage{contour}
\usepackage{url}
\usepackage{cancel}
\usepackage{cite}
\usepackage[colorlinks=true,allcolors=blue,pdfborder={0 0 0},linktocpage=false,pdfencoding=auto]{hyperref}
\usepackage{tabularx,booktabs}
\usepackage{multicol}
\usepackage{feynmp}
\usepackage{units}
\usepackage{xspace}
\usepackage[labelfont=bf]{caption}
\usepackage[section]{placeins}
\usepackage{subcaption}
\usepackage{soul}
\usepackage{bbold}
\usepackage{parskip}
\usepackage{float}
\usepackage{tabulary}

\definecolor{darkred}{rgb}{0.6,0,0}
\definecolor{darkpurple}{rgb}{0.5,0,0.5}

\newcommand{\beqn}{\begin{eqnarray}}
\newcommand{\eeqn}{\end{eqnarray}}
\def\non{\nonumber\\}
\def\se{Sommerfeld ~enhancement~}
\def\s{{\mathsf{s}}}

\begin{document}
\author{
Jinzheng Li$^a$\footnote{\href{mailto:li.jinzh@northeastern.edu}{li.jinzh@northeastern.edu}} ~and
Pran Nath$^a$\footnote{\href{mailto:p.nath@northeastern.edu}{p.nath@northeastern.edu}}\\~\\
$^{a}$\textit{\normalsize Department of Physics, Northeastern University, Boston, MA 02115-5000, USA} \\
}

\title{\vspace{-2cm}\begin{flushright}
\end{flushright}
\vspace{1cm}
\Large \bf
Big Bang initial conditions and self-interacting 
 hidden dark matter
 \vspace{0.0cm}
 }
 
\date{}
\maketitle
\begin{abstract}
A variety of supergravity and string models involve hidden sectors 
where the hidden sectors may couple feebly with the visible sectors via a variety of portals. While the coupling of the hidden sector to the visible
sector is feeble its coupling to the inflaton is largely unknown. It could
couple feebly or with the same strength as the visible sector which would
result in either a cold or a hot hidden sector at the end of reheating. 
These two possibilities could lead to significantly different outcomes
 for observables.  We investigate the thermal evolution of the two sectors in a cosmologically consistent hidden sector dark matter model where the hidden sector and the visible sector are thermally coupled.
  Within this framework we  analyze several phenomena to illustrate their dependence on the initial conditions. These include the allowed parameter space of models, dark matter relic density, proton-dark matter cross section, effective massless neutrino species at BBN time, 
 self-interacting dark matter cross-section, where self-interaction occurs
 via exchange of dark photon, and Sommerfeld enhancement.
Finally fits to the velocity dependence of dark matter cross sections from galaxy scales to the scale of galaxy clusters is given. The analysis indicates
significant effects of the initial conditions on the observables listed above. The analysis is carried out within the framework where dark matter is constituted of dark fermions and the mediation between the visible and  the hidden sector occurs via the exchange of dark photons. The techniques discussed here may have applications for a wider class of hidden sector models using different mediations between the visible and the hidden sectors to explore the 
impact of Big Bang initial conditions on observable physics. 

    \end{abstract}

\numberwithin{equation}{section}
\newpage

{  \hrule height 0.4mm \hypersetup{colorlinks=black,linktocpage=true} \tableofcontents
\vspace{0.5cm}
 \hrule height 0.4mm} 
 
\section{Introduction\label{sec1}}
 Hidden sectors appear in most modern models of particle physics 
beyond the standard model and have become increasingly relevant 
in analyses of particle physics phenomena. Success of precision
electroweak physics 
tell us that the hidden sector couplings to the standard model
must be feeble. But what about the coupling of the hidden sector to 
the inflaton? If the coupling of the hidden sector to the
inflaton is also feeble relative to the coupling of the standard model
 the population of the hidden sector particles
would be negligible and their temperature would be much colder than
 of the standard model particles. On the other 
extreme the hidden sector and the visible sectors could couple
democratically, i.e.,with equal strength, to the inflaton  and 
thus be essentially in thermal equilibrium at the end of reheating.
These cases represent two extreme possibilities with a variety of 
 other possibilities in between. Because
 of the interactions between the visible and the hidden sectors,
 the two sectors are thermally coupled and thus their evolution
 is constrained by the initial condition on the hidden sector 
 at the end of inflation which can be codified by the ratio 
 $\xi_0\equiv T_h^0/T^0$, where $T_h^0$ is the temperature of
 the hidden sector  and $T^0$ is the temperature of the
 visible sector initially after reheating. 
 It is thus of relevance to ask the influence of the initial conditions on
 physical observables at low energy. 
  In this work we study the effect of $\xi_0$ on a variety of physical
  observables, i.e., on the relic density of dark matter, on the proton-DM 
  scattering cross-sections, on the number of massless degrees of freedom
  at BBN, and on DM self-interaction cross-sections. For DM self-interaction 
  cross-section, we further analyze the effect of $\xi_0$ on its velocity
  dependence and on Sommerfeld enhancement and analyze the
  effect of $\xi_0$ on fits to the galactic dark matter cross sections 
   from the scale of dwarf galaxies to the scale
  of galaxy clusters. 
  The portal we utilize in the analysis consists of a hidden sector with a $U(1)_X$ gauge invariance with kinetic mixing\cite{Holdom:1985ag} and 
  Stueckelberg mass growth of the $U(1)_X$ gauge boson~\cite{Kors:2004dx}.  
   By numerically solving the Schrödinger equation, we are able to achieve a comprehensive understanding of the dark matter self-interacting cross-section in this model. While our analysis is done in a specific choice of the portal
  one may expect similar effects discussed in this work using other portals connecting visible and the hidden sectors.
    
  The outline of the rest of the note is as follows:  Section~\ref{sec2}
 gives a summary of the thermal evolution  of coupled visible and hidden
 sectors while section \ref{sec3} discusses a specific model with the 
 visible sector coupled to one hidden sector where  the coupling arises
 via kinetic mixing along with 
 Stueckelberg mass generation for the hidden sector gauge boson. Section 4 discusses the effect of 
 $\xi_0$ on dark freeze-out and on the relic density of dark matter.
 Here we also discuss the dependence of $\Delta N_{\rm eff}$ at BBN time on
 $\xi_0$ and further the effect of $\xi_0$ on the allowed parameter space
 and on spin-independent proton-DM cross section. 
 In section ~\ref{sec5} we discuss the effect of $\xi_0$ on Sommerfeld enhancement  for the self-interacting dark matter cross section. In section \ref{sec6} we discuss the effect of $\xi_0$ on fits to the galaxy data
 on dark matter cross sections  from  low relative velocities to high relative velocities which encompass scales from dwarf 
 galaxies to galaxy clusters. Conclusions are given in section ~\ref{sec7}.    
  Additional details related to the analysis are given in sections 
 \ref{appen:A}-\ref{appen:C}. Further, in section \ref{appen:D} we give 
 a comparison of our analysis of the thermal evolution when the total entropy is conserved
  vs the thermal evolution when the entropies of the visible and the hidden
 sectors are separately conserved.
 In this section we also analyze the accuracy of using conservation of total entropy
 in computations of yields for dark matter since in general the total entropy is not conserved unless the sectors equilibrate.

\section{Cosmologically consistent evolution of coupled visible and hidden sectors in DM analysis.}\label{sec2}
As mentioned in the previous section,
most models of particle physics based on physics beyond the standard 
model contain hidden sectors which may be feebly coupled to the 
visible sector. In this case the thermal evolution of each is interdependent
on the other. Thus the approximation typically made that the entropy of 
the visible and the hidden sectors are separately conserved is invalid.
{Further,} the hidden sector by itself may consist of several
sectors some of which may interact directly with the visible sector 
while others indirectly via their interactions with other hidden sectors
which couple with the visible sector. First, in this case  the
Hubble expansion is affected by the hidden sectors via their energy
densities so that
\begin{align}
H^2&= \frac{8\pi G_N}{3} (\rho_v + \sum_{i=1}^n \rho_i)\,,
\label{hubble}
\end{align}
where $\rho_v$ is the energy density of the visible sector and $\rho_i$
the energy density of the $i$-th hidden sector where $\rho's$ have temperature
dependence so that 
\begin{align}
\rho&=\frac{\pi^2}{30}\left(g_{\rm eff}^v T^4+\sum_{i=1}^ng^h_{i\rm eff} T_i^4
\right).
\label{rho-s}
\end{align}
and the  
  total entropy density of the visible and hidden sectors is given by
\begin{align}
\s&=\frac{2\pi^2}{45}\left(h_{\rm eff}^v T^3+\sum_{i=1}^n h^h_{i\rm eff} T_i^3
\right).
\label{entropy}
\end{align}
Here $g^{v,h}_{\rm eff}$ and $h^{v,h}_{\rm eff}$ are the energy and entropy degrees of freedom and are temperature dependent.
A full expression for them for the specific model we will consider is given in 
section 3. 
In \cite{Aboubrahim:2022bzk} an analysis was given where 
  the visible sector (V) at temperature $T$  is coupled to the
hidden sector $H_1$ at temperature $T_1$, the hidden sector $H_1$ is coupled to the hidden sector $H_2$ at temperature $T_2$,  and so on and finally that
the hidden sector $H_{n-1}$ at temperature $T_{n-1}$ is coupled to the hidden sector $H_n$ at temperature $T_n$. In the analysis of \cite{Aboubrahim:2022bzk} radiation dominance was assumed. Here we
extend the analysis to include radiation and matter.
In this case the energy densities for various
sectors obey the following set of coupled Boltzmann equations
\beqn
\frac{d\rho_\alpha}{dt} + 3 H(\rho_\alpha+p_\alpha)&=j_\alpha,~~\alpha=0,1,2,\dots,n\,,
\label{2.1} 
\eeqn
Here $\rho_\alpha$ and $p_\alpha$ are the energy and momentum densities
for the sector $\alpha$, and 
where  $\alpha=0$ refers to the visible sector, and $\alpha=1,2,\cdots,n$ to the 
 hidden sectors,
  and where $j_\alpha$ encodes in it all the possible processes exchanging energy between neighboring sectors. 
We note now that the total energy density $\rho=\sum_{\alpha=0}^{n}
\rho_\alpha$ in an expanding universe satisfies the   equation
\begin{align} 
\frac{d\rho}{dt}+ 3H(\rho+p)=0.
\end{align}
where $p=\sum_{\alpha=0}^n \rho_\alpha$ is the total pressure density.
In the analysis it is convenient to  introduce the functions  
$\zeta = \frac{3}{4}(1+\frac{p}{\rho})$ and  $\zeta_\alpha = \frac{3}{4}(1+\frac{p_\alpha}{\rho_\alpha})$, where $\zeta_\alpha=1$ for radiation 
dominance and $\zeta_\alpha=\frac{3}{4}$ for matter dominance. 
  More generally $\zeta$ and $\zeta_\alpha$ are temperature dependent
  and this dependence is taken into account in the evolution equations.
  Thus $\rho_\alpha$ and $\rho$ satisfy the  evolution equations 

\begin{align} 
\label{1.26}
\frac{d\rho_\alpha}{dt}+ 4H\zeta_\alpha\rho_\alpha=j_{\alpha},\\
\frac{d\rho}{dt}+ 4H\zeta\rho=0.
\label{1.27}
\end{align}
We use the visible sector temperature $T$ as the clock as thus wish to 
write the evolution equations  Eqs. (\ref{1.26}) and (\ref{1.27}) in terms of temperature $T$ rather than time. This is accomplished using the relation
\begin{align} 
\frac{dT}{dt} = -\frac{4H\zeta\rho}{\frac{d\rho}{dT}}.
\end{align}
Thus using  $d\rho_\alpha/dt=(d\rho_\alpha/dT)(dT/dt)$, one has 
 \beqn
 \frac{d\rho_\alpha}{dT}=\frac{(4H\zeta_\alpha\rho_\alpha-j_\alpha)}{4H\zeta\rho} \frac{d\rho}{dT}.
 \label{dr-dT}
 \eeqn
 Next decomposing $\rho$ so that $\rho=\rho_v+ \sum_{i=1}^n \rho_i$,
one finds that  $d\rho_i/dT$ can be written as 
\beqn
\frac{d\rho_i}{dT}=\sum_{j=1}^n ({\cal C}^{-1})_{ij} C_j \frac{d\rho_v}{dT}.
\label{dridT}
\eeqn
Here
\begin{equation}
C_i=\frac{4H\zeta_i\rho_i-j_i}{4H\sigma_i+j_i},   
\end{equation}
 where $\sigma_i= \zeta\rho-\zeta_i\rho_i$ and ${\cal C}_{ij}$ is defined so that 
\begin{equation}
 {\cal C}_{ij}= \delta_{ij}- C_i (i\neq j), i, j=1,2,\cdots, n. 
\end{equation}
Note that one may also write 
\beqn
&\frac{d\rho_i}{dT}= P_i + Q_i \xi_i', ~~~i=1,2,\cdots,n.
\label{2.} 
\eeqn
Here $P_i= \xi_i \frac{d\rho_i}{dT_i},~Q_i= T \frac{d\rho_i}{dT_i}$, 
where $\xi_i= T_i/T$ and  $ \xi'_i\equiv d\xi_i/dT$.
Thus we have an equation for $d\xi_i/dT$ which takes the form
\beqn
\frac{d\xi_i}{dT}=-\frac{P_i}{Q_i} +     \sum_{j=1}^n({\cal C}^{-1})_{ij} C_j\frac{\rho_v'}{Q_i};  ~~i=1,2,\cdots,n.
\label{2.a10} 
\eeqn
where $\rho_v'=d\rho_v/dT$.
Eqs.~(\ref{2.a10}) give us a set of $n$ differential equations for the evolution functions $d\xi_i/dT$. These have to be solved along with the Boltzmann equations governing the number density evolution of the hidden sector particles. This will allow us to determine the relic densities of all stable species and describe the thermal evolution of this coupled system. 
For the case of the visible sector coupled to one hidden sector we have 
$C_{11}=1, C_1=(4H\zeta_h\rho_h-j_h)/(4H\zeta\rho-4H\zeta_h\rho_h+ j_h)$, $\rho_h\equiv \rho_1$, $T_h\equiv T_1$, $j_h\equiv j_1$, and we define
$\xi\equiv \xi_1=T_h/T$. The source term $j_h$ is discussed in 
section \ref{appen:A}.
 With this notation specific to the case of the visible sector and 
one hidden sector we have the following equation for $\xi$ which governs the
temperature evolution of the hidden sector relative to that of  the visible sector
\beqn
\frac{d\xi}{dT}= \left[ -\xi \frac{d\rho_h}{dT_h} +
\frac{4H\zeta_h\rho_h-j_h}{4H\zeta\rho-4H\zeta_h\rho_h+ j_h} \frac{d\rho_v}{dT}\right] (T \frac{d\rho_h}{dT_h})^{-1}.
\label{DE1}
\eeqn
We note that $g^v_{\rm eff}$ and $h^v_{\rm eff}$ are pre-calculated and we use tabulated results from micrOMEGAs \cite{Belanger:2018ccd}.  
As noted already
 $g^h_{\rm eff}$ and $h^h_{\rm eff}$ for the hidden sector 
 that enter Eq. (\ref{rho-s}) and Eq. (\ref{entropy})  
 are temperature dependent
 \cite{Hindmarsh:2005ix,Husdal:2016haj} and their explicit expressions are given in  Eq.(\ref{geff-heff}). 
\section{{The model coupling visible and hidden sectors}} \label{sec3}
There are a variety of portals that allow communication between the visible and the
hidden sectors. These include the Higgs field portal~\cite{Patt:2006fw}, kinetic mixing
of two gauge fields ~\cite{Holdom:1985ag}, Stueckelberg mass mixing \cite{Kors:2004dx,Kors:2005uz}, kinetic and Stueckelberg mass mixing~\cite{Feldman:2007wj},
Higgs-Stueckelberg portal\cite{Du:2022fqv}, as well as other possibilities such as higher dimensional operators.
In this work we focus on kinetic mixing along with
the mass growth for the hidden sector gauge field via the Stueckelberg mechanism.
Thus for analysis in this work we consider a specific model 
for dark matter which
is an extension of the standard model with an $SU(3)\times SU(2)\times U(1)_Y\times U(1)_X$ gauge invariance where the  $U(1)_X$ gauge field has
kinetic mixing with the visible sector $U(1)_Y$  gauge field~\cite{Holdom:1985ag} 
and 
in general a Stueckelberg mass mixings
~\cite{Kors:2004dx,Feldman:2007wj, Aboubrahim:2020lnr,Aboubrahim:2019qpc}.
We assume that the $U(1)_X$ hidden sector has a dark fermion $D$
which interacts with the $U(1)_X$ gauge field. Thus the 
 extended $SU(3)\times SU(2)\times U(1)_Y\times U(1)_X$ Lagrangian
 consisting of the SM part $\mathcal{L}_{SM}$ 
 and  the extended part  $\mathcal {L}_{\text{ext}}$  
 is given by 
 \begin{align}
 \mathcal {L}&= \mathcal{L}_{SM} + \mathcal{L}_{\text{ext}}\,,\nonumber\\
 \mathcal{L}_{\text{ext}}&= -\frac{1}{4} C^{\mu\nu} C_{\mu\nu} 
 -\bar D (\gamma^\mu \frac{1}{i} \partial_\mu + m_D) D
 - g_X \bar D Q_X\gamma^\mu DC_\mu 
 \nonumber\\
 &~~~~
 - \frac{\delta}{2} C^{\mu\nu} B_{\mu\nu} - \frac{1}{2}(M_1C_\mu + M_2 B_\mu + \partial_\mu \sigma)^2,
 \label{basic-lag}
 \end{align}
 Here $B_\mu$ is the gauge field for the $U(1)_Y$, 
 $C_\mu$ is the gauge field of $U(1)_X$,  
 $\sigma$ is an axionic field which gives mass to $C_\mu$ and is absorbed in the unitary gauge and $D$ is the dark fermion where $Q_X$ is the $U(1)_X$
 charge of $D$ and $g_X$ is gauge coupling of $U(1)_X$.
 Further, $\delta$ is the kinetic
 mixing parameter between the field strengths of $C^\mu$ and $B^\mu$, 
 and  $M_1$ and $M_2$  are the Stueckelberg 
 mass parameters. A non-vanishing $M_2$ will lead to a milli-charge for the
 dark fermion $D$ and we assume neutrality of dark matter and thus set 
 $M_2=0$ in the analysis\footnote{A non-vanishing $M_2$ was used 
 to resolve the EDGES anomaly in the analysis of \cite{Aboubrahim:2021ohe}.}.
The spontaneous breaking of the $SU(2)\times U(1)_Y$ 
 electroweak symmetry along with the Stueckelberg mass growth
 gives rise to mixing among the three gauge fields $C^\mu, B^\mu, A^\mu_3$
 where $A^\mu_3$ is the third component of the $SU(2)_L$ gauge field
 $A^\mu_a (a=1,2,3)$ of the standard model.  The mixings give rise to a $3\times 3$ mass square matrix which can be diagonalized by the three Euler angles $(\theta, \phi, \psi)$ which are given in Eq.(\ref{angles}). 
 The diagonalization gives the following mass eigenstates:  the $Z$ boson, a massive dark photon $\gamma'$ and the massless
 photon $\gamma$.
  The Lagrangian governing the interaction of 
dark photon and dark fermion
 which enters into our analysis is given by
\begin{align}
\mathcal{L_{\text{dark}}}=- \frac{1}{4} A_{\mu\nu \gamma'} 
A_{\gamma'}^{\mu\nu} -\frac{1}{2} m^2_{\gamma'} 
A_{\gamma'\mu}A_{\gamma'}^\mu 
 -\bar D (\gamma^\mu \frac{1}{i} \partial_\mu + m_D) D
- \bar{D}\gamma^{\mu}\left[\epsilon^D_Z Z_{\mu}+g_{\gamma'}^D A^{\gamma'}_{\mu}\right]D.
 \label{ddgammap}
 \end{align}

 The interaction of Eq. \ref{ddgammap} involves two massive gauge 
 bosons ($Z, \gamma'$).  For the case when the kinetic mixing
 is small one  has  $g_{\gamma'}^D\simeq   g_XQ_X$ and $\epsilon^D_Z 
 =O(\delta^3)$ which is negligible.
 In addition the dark photon will have couplings 
 with the standard model quarks and leptons which
 are discussed in Appendix 10.
 Setting $Q_X=1$, the input parameters of the model
 are $g_X, m_D, m_{\gamma'}, \delta$ which are 
 what appear in Table 1. We note here that models with the vector boson as the mediator between the hidden sector 
 and the visible sector have been considered in several
 previous works 
~\cite{Chen:2006ni,Feldman:2006wd,Feldman:2007nf,Cheung:2007ut,Feng:2012jn,Feldman:2011ms,Feng:2013wn, Chen:2009iua,Buckley:2009in,Ibe:2009mk,Feng:2009mn,Loeb:2010gj,Tulin:2012wi,Tulin:2013teo,Schutz:2014nka,Feng:2008mu,Redondo:2008ec,Chu:2011be,Chu:2013jja,Kaplinghat:2015aga,DAgnolo:2015ujb,Kuflik:2015isi,Bringmann:2016din,Evans:2019vxr,Sagunski:2020spe,Fernandez:2021iti,Kaneta:2016wvf,Kaneta:2017wfh,Co:2018lka,Dror:2018pdh,Agrawal:2018vin,Long:2019lwl,AlonsoAlvarez:2019cgw,Nakai:2020cfw,Choi:2020dec,Delaunay:2020vdb,Graham:2015rva,Ema:2019yrd,Ahmed:2020fhc,Berger:2016vxi,Ibe:2019gpv,McDermott:2019lch,Fradette:2015nna}. 
   Axions and dark photons in the light to 
 ultralight mass region  have also been investigated
~\cite{Long:2019lwl,Bloch:2016sjj,Pospelov:2008jk,Nakai:2020cfw,Kim:2015yna,Hui:2016ltb,Halverson:2017deq}, and dark photons have been used in explaining astro-physical phenomena including galactic $\gamma$-rays
 \cite{Boehm:2003bt,Boehm:2003hm} and 
 PAMELA  positron excess \cite{Feldman:2008xs,ArkaniHamed:2008qn,Pospelov:2007mp,Bergstrom:2013jra,Bergstrom:2008gr}.

We further note that the dark photon in this model even when
very light and kinematically disallowed to decay into $e^+e^-$ will 
eventually decay via the modes $\gamma'\to \nu\bar \nu$ and 
$\gamma'\to 3 \gamma$ and not contribute to dark matter density unless its 
lifetime is larger than the lifetime of the universe and even in  that case only if
it has non-negligible relic density, which is not the case we consider.
Thus the dark fermion will be the only constituent of dark matter. Further details about this model is given in Appendix \ref{appen:Modeldetial}.
  \section{Big Bang constraints on dark freeze-out,
    relic density, $\Delta N_{\text{\rm eff}}$, and on proton-DM cross-section}   \label{sec4} 
  In this section we discuss the effects on the relic density,  on the number of 
  relativistic  
  degrees of freedom due to the hidden sector at the BBN time,
  on the allowed parameter space of models, and on  proton-dark matter scattering scattering cross section arising
  from different choices of the initial value $\xi_0$  at the end of reheating.
   In the model discussed in the preceding section, 
 the dark fermion $D$ constitutes
  dark matter and has self-interactions due to exchange of dark photon. 
  \subsection{Effect of $\xi_0$ on dark freeze-out and on relic density}
  In the analysis here we will discuss the effect of $\xi_0$ on the
  dark freeze-out which generates the relic density of $D$. 
  Computationally the quantities of  interest for this purpose 
  are the yields for the dark fermion  $Y_D$ and for the dark
photon $Y_{\gamma'}$, where the yield is defined so that 
$Y=n/\s$ where $n$ is the number density and $\s$ is the entropy
density. Assuming conservation of total entropy (this assumption will be tested in section \ref{appen:D}) the evolution equations for  $Y_D$ and $Y_{\gamma'}$  
are given by

\begin{align}
    \frac{dY_D}{dT}  =& -\frac{\s}{H}\left( \frac{d\rho_v/dT}{{4\zeta\rho-4\zeta_h\rho_h}+j_h/H}\right) \Big[\left<\sigma v\right>_{D\bar{D} \rightarrow i\bar{i} }(T)Y_D^{eq}(T)^2 -\left<\sigma v\right>_{D\bar{D} \rightarrow \gamma'\bar{\gamma'} }(T_h)Y_D(T_h)^2\non
   &  +\left<\sigma v\right>_{\gamma'\bar{\gamma'} \rightarrow D\bar{D} }(T_h)Y_{\gamma'}(T_h)^2 \Big]\,,\label{DE2}\\
    \frac{dY_\gamma'}{dT}  =& -\frac{\s}{H}\left( \frac{d\rho_v/dT}{{4\zeta\rho-4\zeta_h\rho_h}+j_h/H}\right)  \left[\left<\sigma v\right>_{D\bar{D} \rightarrow \gamma'\bar{\gamma'} }(T_h)Y_D(T_h)^2 -\left<\sigma v\right>_{\gamma'\bar{\gamma'} \rightarrow D\bar{D} }(T_h)Y_{\gamma'}(T_h)^2 +\right.\non
    &\left.\left<\sigma v\right>_{i\bar{i}\rightarrow \gamma' }(T)Y_{i}^{eq}(T)^2 -\left<\Gamma_{\gamma'\rightarrow i\bar{i}(T_h)}\right>Y_{\gamma'}(T_h) \right]\,.\label{DE3}
\end{align}

Here $\left<\sigma v\right>_{D\bar{D} \rightarrow i\bar{i}}$ is the
annihilation cross-section of $D\bar D$ into standard model particles
which are denoted by $i\bar i$,  $\left<\sigma v\right>_{D\bar{D} \rightarrow \gamma'\gamma'}$ is their annihilation into dark photon 
while $\left<\sigma v\right>_{i\bar{i} \rightarrow \gamma'}$ gives
 the annihilation of standard model particles into a dark photon,
and $n_D$ and $n_{\gamma'}$ are the number densities of the $D$ fermion
 and the dark photon $\gamma'$.
In the above the cross-section for the process $D\bar D\to \gamma' \gamma'$ is given by 
\begin{align}
&\sigma^{D\bar{D}\to {\gamma'} {\gamma'}}(s)=\frac{g_{X}^4(\mathcal{R}_{11}-s_{\delta}\mathcal{R}_{21})^4}{8\pi s(s-4m^2_D)}
\times\Bigg\{-\frac{\sqrt{(s-4m^2_{{\gamma'}})(s-4m^2_{D})}}{m^4_{{\gamma'}}+m^2_D(s-4m^2_{{\gamma'}})}[2m^4_{{\gamma'}}+m^2_D(s+4m^2_D)] \nonumber \\
&+
log\Big[\frac{s-2m^2_{{\gamma'}}+\sqrt{(s-4m^2_{{\gamma'}})(s-4m^2_{D})}}{s-2m^2_{{\gamma'}}-\sqrt{(s-4m^2_{{\gamma'}})(s-4m^2_{D})}}\Big]
\frac{(s^2+4m^2_D s+4m^4_{{\gamma'}}-8m_D^4-8m_D^2 m^2_{{\gamma'}})}{s-2m^2_{{\gamma'}}}
\Bigg\}.
\label{DDgg}
\end{align}
 while the rest of the cross-sections are given in appendix of \cite{Aboubrahim:2021ohe}.
Here $s,t,u$ are Mandelstam variables where $s+t+u= 2 m_D^2
+2m_{\gamma'}^2$, 
$\mathcal{R}_{11}$ and $\mathcal{R}_{21}$ are matrix elements of $\mathcal{R}$ 
which diagonalizes the mass and kinetic 
energy matrices of Eq.(\ref{basic-lag})
as given in~\cite{Feldman:2007wj}. Further, we note that 
 in addition to $D\bar D\to \gamma'\gamma'$
we also have $\gamma'\gamma'\to D\bar D$ which enters in the yield
equations when kinematically allowed and is related to $D\bar D\to \gamma'\gamma'$ so that 
\begin{equation} 
9(s-4m_{\gamma'}^2)\sigma^{\gamma'\gamma'\to D\bar D}(s) = 8(s-4m_D^2)\sigma^{D\bar{D}\to {\gamma'} {\gamma'}}(s).
\end{equation} 
In the above equation, the thermally averaged cross section and decay widths are given by
\begin{align}
\label{k1k2a}
  \left<\sigma v\right>^{a\bar{a}\rightarrow bc}(T) = &\frac{1}{8m_a^4TK_2^2(m_a/T)}\int_{4m_a^2}^{\infty}ds\sigma(s)\sqrt{s}(s-4m_a^2)K_1(\sqrt{s}/T).\\
  \left<\Gamma_{X\rightarrow i\bar{i}}(T)\right>=&\Gamma_{x\rightarrow i\bar{i}}\frac{K_1(m_X/T)}{K_2(m_X/T)}.
  \label{k1k2b}
\end{align}
The equilibrium yield of the i-th particle 
 is given by
\begin{align}
    Y_i^{eq} = \frac{n^{eq}_i}{\s} = \frac{g_i}{2\pi \s}m_i^2 T K_2(m_i/T)\,,
    \label{k1k2c}
\end{align}
where $\s$ is the entropy density.  In Eqs.(\ref{k1k2a}), (\ref{k1k2b}) and (\ref{k1k2c}), $K_1$ and $K_2$ are the modified Bessel functions of the second kind and of degrees one and of degree two. Further, cross sections for the processes
 $D\bar D \to i \bar i$, where $i,\bar i$ are the standard model particles,
 can be found in Appendix D of \cite{Aboubrahim:2021ycj}.  
As noted above in this model the dark photon is unstable and decays and does not
contribute to the relic density and the entire DM relic density arises
from the dark fermion where at current times the relic density 
$\Omega_{D}h^2$ is given by 
\begin{align}
    \Omega_D h^2 = \frac{\s_0m_D Y^0_D h^2}{\rho_c}\,,
    \label{relic}
\end{align}
where $\s_0$ is the current entropy density, $Y^0_D$ which is $Y_D$ 
at current times can be gotten using  Eqs. (\ref{DE1}), (\ref{DE2}) and (\ref{DE3}), $\rho_c$ is the critical energy density needed to 
close the universe, and  $h$ is defined so that $H_0=100 h$ km s$^{-1}$
Mpc$^{-1}$, where $H_0$ is the Hubble parameter today.

The procedure for solving the evolution equations involves simultaneous analysis of coupled equations Eq.(\ref{hubble})-Eq.(\ref{entropy}),
  Eq.(\ref{DE1}),  
  Eq.(\ref{geff-heff}),
 Eq.(\ref{rhoh-ph}), the 
 yield equations for $Y_D$, $Y_\gamma'$,  Eq.(\ref{DE2})-Eq.(\ref{DE3}),
 and Eqs.(\ref{DDgg})-(\ref{relic}). {{Using these we
 do Monte Carlo simulations with parameters varying  in the ranges} 
 \begin{align}
     10^{-1} \rm{GeV} < m_{D}< 10^3 \rm{GeV},&\quad 10^{-2} \rm{MeV} < m_{\gamma'}< 10^2 \rm{MeV}\nonumber\\
     \quad 10^{-4}  < g_X< 1 ,&\quad  10^{-12}  <\delta< 10^{-7}.
 \end{align}
 and search for model points satisfying all the current experimental constraints.}
Table \ref{tab:benchmarks} gives 6 model points used in this paper
all of which are consistent with the current experimental constraints
\cite{Aboubrahim:2022qln}  including those from a variety of experiments, i.e., 
 BaBar~\cite{BaBar:2014zli,BaBar:2017tiz}, HPS~\cite{HPS:2016jta}, LHCb~\cite{LHCb:2019vmc}, Belle-2~\cite{Belle-II:2018jsg}, SHiP~\cite{SHiP:2020vbd}, SeaQuest~\cite{Berlin:2018pwi,Tsai:2019buq} and NA62~\cite{Tsai:2019buq},
CHARM~\cite{Tsai:2019buq}, $\nu$Cal~\cite{Tsai:2019buq,Blumlein:2011mv,Blumlein:2013cua}, E137~\cite{Andreas:2012mt}, E141~\cite{Riordan:1987aw}, NA64~\cite{NA64:2019auh}, NA48~\cite{NA482:2015wmo}. For a sub-MeV dark photon mass stringent constraints arise from Supernova SN1987A~\cite{Chang:2016ntp} and from BBN, stellar cooling~\cite{An:2013yfc} and from the decay to $3\gamma$ on cosmological timescales~\cite{Essig:2013goa,Redondo:2008ec}
 An analysis of these constraints in limiting the parameter space have been analyzed in\cite{Aboubrahim:2022qln,Aboubrahim:2021ycj}. The parameter space chosen in the 
 current analysis is consistent with these constraints. In fact, a larger variation in \\
 \begin{table}[h]
 \centering
     \begin{tabular}{llllll}
    \hline
   No. &  $m_D[\text{GeV}]$& $m_{\gamma'}[\text{MeV}]$&$g_X$ & $\delta$(in $10^{-9}$)  & $\tau_{\gamma'\to 3 \gamma}$(yrs)\\ \hline
(a)&0.354&0.306&0.00738&3.99& $2.6\times 10^7$\\
(b)&0.259&0.214&0.00675&6.29&$2.6\times 10^8$  \\(c)&0.281&0.550&0.00931&400& $1.3\times 10^1$  \\
(d)&0.170&0.225&0.00618&19.3&  $1.8\times 10^7$    \\(e)&0.156&0.285&0.00631&52.9& $2.8\times 10^5$    \\(f)&0.568&0.445&0.00810&2.62 &  $2.0\times 10^6$    \\
   \end{tabular}
    \caption{6 model points used in the analysis of this work and their decay lifetime for the dark photon.}
    \label{tab:benchmarks}
\end{table}

We note here that the mass of the dark photon is in the sub MeV region and is 
long lived with its most dominant decay mode being $\gamma'\to 3 \gamma$. 
For kinetic mixing the decay width for the mode $\gamma'\to 3 \gamma$ is given by~
\cite{Redondo:2008ec,Essig:2013goa,McDermott:2017qcg}
\begin{align}
 \Gamma_{\gamma'\to 3 \gamma}= \frac{17 \alpha^4 (\epsilon^{\gamma}_{\gamma'})^2}{ 2^7 3^65^3\pi^3} \frac{m_{\gamma'}^9}{m_e^8}.
\end{align}	
where $\alpha=\frac{e^2}{4\pi}$, $\epsilon^{\gamma}_{\gamma'}$ is the kinetic mixing parameter of coupling between dark photon $\gamma'$ and photon $\gamma$ given by $\epsilon^{\gamma}_{\gamma'} =  g_Y \sqrt{1+\bar{\epsilon}^2}\mathcal{R}_{21}$ 
as defined in \cite{Feldman:2007wj},
 $m_{\gamma'}$ is the
dark photon mass and $m_e$ is the electron mass.
The dark photon lifetimes for the different model points are given in Table (\ref{tab:benchmarks}). Here we find that the dark photon lifetimes are smaller than the age of the universe and 
thus there is no contribution of the dark photon to the relic density and consequently no constraint on the 
allowed parameter space regarding the  relic density constraint. We note, however, that even if the dark photon was long lived with a lifetime greater than the lifetime of the universe, its contribution 
to the relic density would be negligible.  A 
recent analysis~\cite{Aboubrahim:2021ycj} 
in accord with the analysis of \cite{Redondo:2008ec} shows  that with one hidden sector it is not possible to get both a long lived dark photon which can contribute to the relic density and simultaneously achieve a significant amount of dark matter relic density. To do that one needs at least a two hidden sector model\cite{Aboubrahim:2021ycj} in which
a dark photon as dark matter can produce a non-negligible amount of dark matter. 
  \begin{figure}[h]
   \centering
  \includegraphics[width=0.4\linewidth]{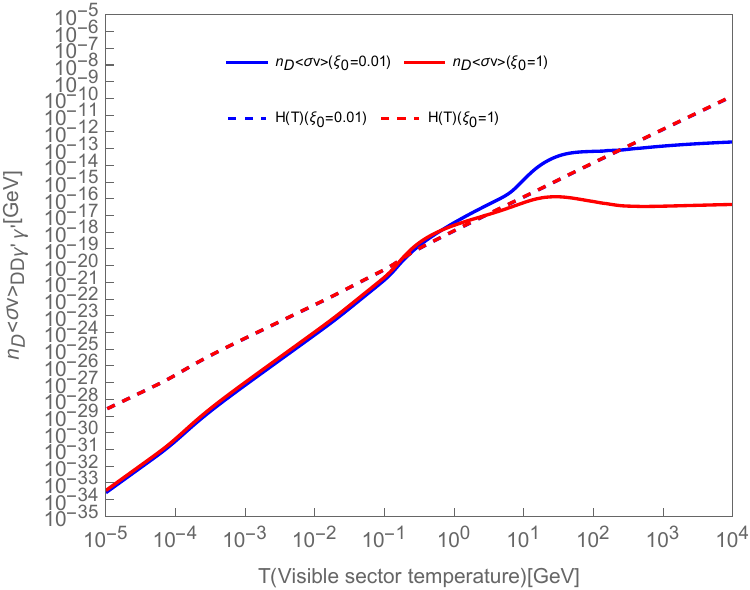} 
    \includegraphics[width=0.4\linewidth]{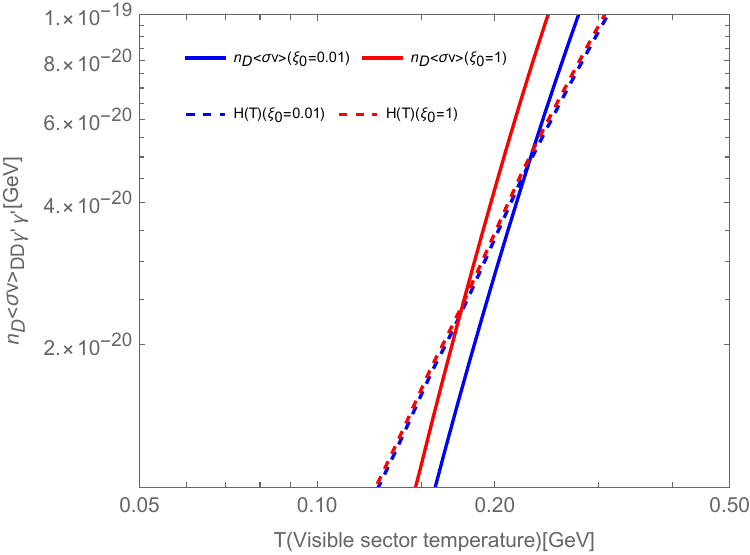} \\
  \includegraphics[width=0.4\linewidth]{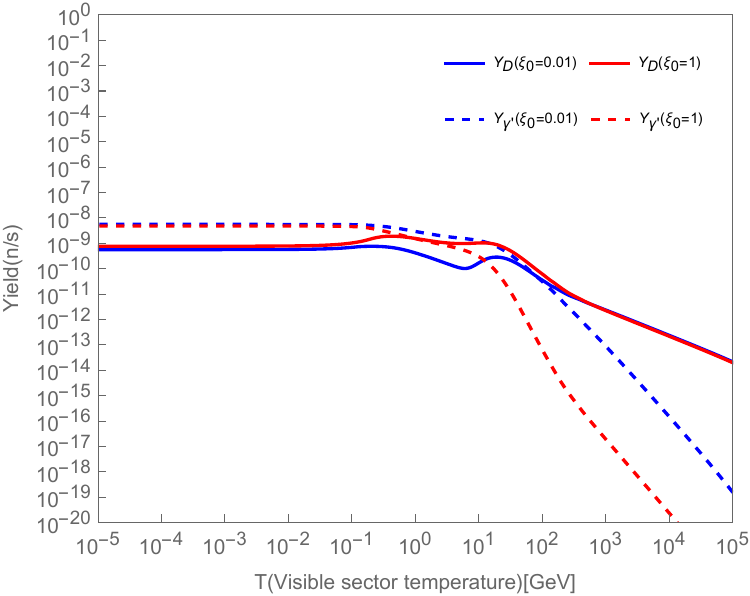} 
  \includegraphics[width=0.4\linewidth]{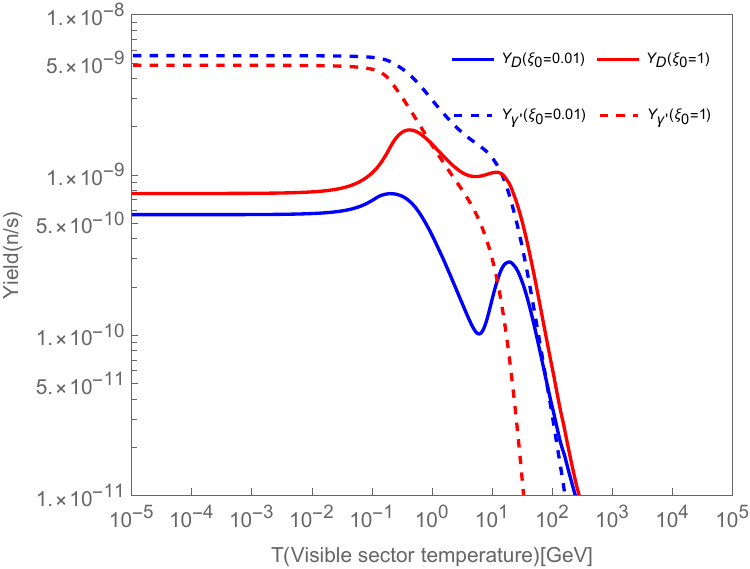}   
\caption{{Top left panel: Exhibition of the dependence
of the dark freeze-out temperature when  $\xi_0=0.01$ (blue) vs $\xi_0=1$ (red)
 for model (f) in Table  \ref{tab:benchmarks}.
Top right panel: zoom in of the top left panel in the region of the 
freeze-out.  Bottom left panel:  Yields of dark fermion (dark matter) and dark photon for  model of the top  panels for $\xi_0=0.01$ (blue), and $\xi_0=1$ (red). Bottom right panel: zoom in of the bottom left panel to exhibit the shift 
of the dark fermion yield for the cases $\xi_0=0.01$ (blue) and $
\xi_0=1$ (red).}}
\label{fig:relicchange}
\end{figure}

   \begin{figure}[h]
   \centering
   \includegraphics[width=0.4\linewidth]{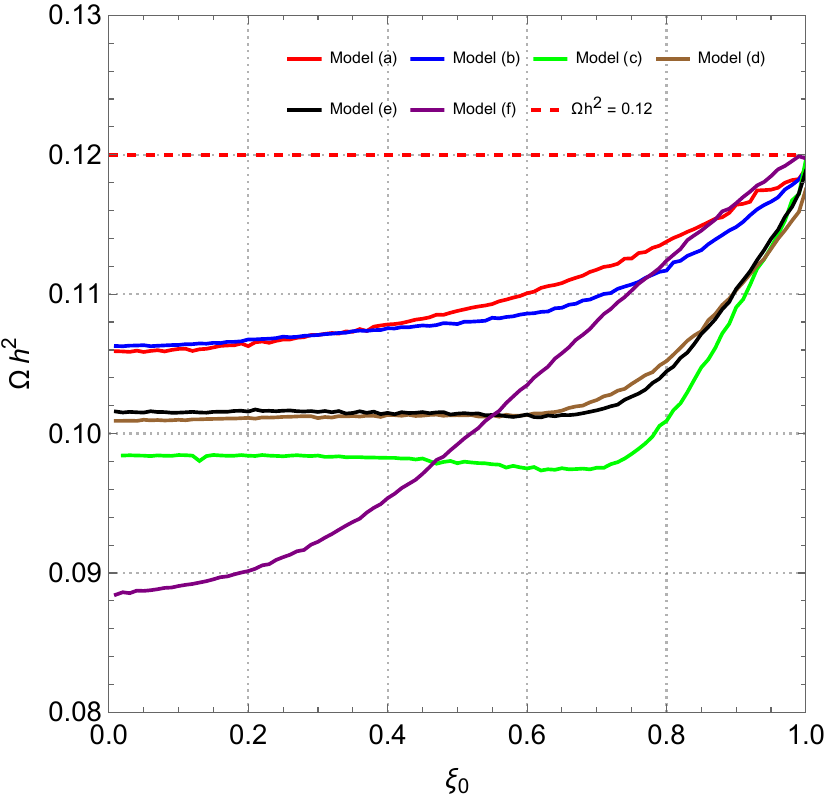}     
           \includegraphics[width=0.4\linewidth]{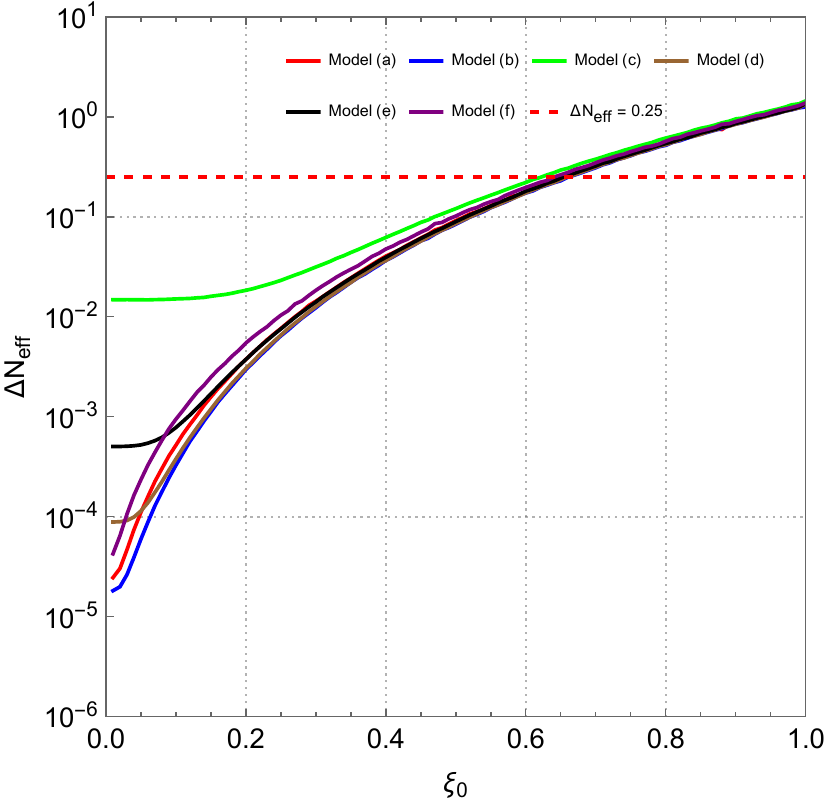}      
\caption{
Left panel: Exhibition of the dependence of the relic 
   density $\Omega h^2$   
   on $\xi_0$ in the range $\xi_0=(0,1)$ for the model points of Table \ref{tab:benchmarks}.
   Right panel:  Exhibition of the dependence of $\Delta N_{\rm eff}$ at BBN
   time on 
   $\xi_0$ in the range $\xi_0=(0,1)$ for the model points of Table \ref{tab:benchmarks}.}
\label{fig-a}
\end{figure}

   We discuss now the dependence of the dark matter freeze-out 
   and of the relic density on the initial conditions.  In the top left panel of 
   Fig. \ref{fig:relicchange}   
   we exhibit  the dependence of the dark freeze-out
   and specifically the decoupling of the dark photon and the dark fermion on 
   $\xi_0$ where we consider the cases: $\xi_0=0.01$ and $\xi_0=1$. 
   The top right panel is the zoom in of the top right in the region of the 
   freeze-out.
   From Fig. \ref{fig:relicchange}   
   we see  that the process $D\bar{D}\rightarrow \gamma'\gamma'$ falls below H(T) at different temperatures for $\xi_0=0.01$
   and for $\xi_0=1$ and consequently the temperature where the dark freeze-out occurs changes by a significant amount.   The sensitivity of the freeze-out on $\xi_0$  directly affects the yields as shown in the bottom left panel and
   the bottom right panel (a zoom in of the bottom left panel) of Fig. \ref{fig:relicchange}.   
  In the left panel of Fig. \ref{fig-a}
we exhibit the dependence of the relic 
   density on $\xi_0$ for the six model points of Table \ref{tab:benchmarks}. 
    Here we find that the relic density can change up to 40\% as $\xi_0$ varies in the range $(0,1)$.
\subsection{Dependence of $\Delta N_{\rm eff}$ at BBN on $\xi_0$}
  One of the predictions of beyond the standard model physics is $N_{\rm eff}$, the number of effective relativistic degrees of freedom at BBN.
   For the standard model  $N_{\rm eff}= 3.046$. The current experimental
   constraint on $N_{\rm eff}$ is summarized in Fig. 39 of the Planck
   Collaboration \cite{Planck:2018vyg} which shows the spread in 
   $N_{\rm eff}$. Thus the Planck Collaboration gives 
   $N_{\rm eff}= 2.99\pm 0.17$ while    the 
   joint BBN analysis of deuterium/helium abundance
    and the Planck CMB data gives $N_{\rm eff}=3.41\pm 0.45$.
    Here we will use the conservative constraint 
    on $\Delta N_{\rm eff}=
    N^{\rm exp}_{\rm eff}- N^{\rm sm}_{\rm eff}$ so that
   $\Delta N_{\rm eff}\leq 0.25$.   
In the model under discussion, the dark fermion D and dark photon $\gamma'$ will contribute to the effective neutrino number. Such contribution 
is given by
\begin{align}
\Delta N_{\rm eff}&=\frac{4}{7} g^h_{\rm eff}\left(\frac{11}{4}\right)^{4/3}
\left(\frac{T_h}{T}\right)^4\,,
\label{neff}
\end{align}
where $g^h_{\rm eff}$ can computed from Eq.(\ref{geff-heff}) and
Eq.(\ref{neff}) is to be evaluated at the BBN temperature
 $T_{\rm BBN}=1$ MeV (for related works see, e.g., 
\cite{Hasegawa:2019jsa,Kawasaki:1999na}).
  In Table \ref{tab:my_label}, $\Delta N_{\rm eff}$ is computed for the six  model points 
 of Table \ref{tab:benchmarks} for $\xi_0=0.01$ and $\xi_0=1$ while 
   the right panel of Fig. \ref{fig-a} exhibits
    $\Delta N_{\rm eff}(\text{\rm BBN})$ for the six model points for $\xi_0$ in the range $(0-1)$.
         The analysis  in general indicates that hidden sectors which start off cooler than the standard model at the end of reheating 
 contribute a smaller amount to  $\Delta N_{\rm eff}$ than those which are relatively hotter at the end of reheating. Further, the analysis indicates that 
 models where $\xi\simeq 0$ could accommodate more massless degrees of freedom
 allowing for the possibility of building a wider class of models
 with more hidden sector particles which may still be consistent with the
   $\Delta N_{\rm eff}$ constraint at BBN time.
  \begin{table}[h]
    \centering
    \begin{tabular}{l|ll|ll}
        \hline
       Model  & \multicolumn{2}{c|}{$\xi_0=1$} & \multicolumn{2}{c}{$\xi_0=0.01$}  \\
              & $\Delta N_{\rm eff}$& $\xi(T_{\rm BBN})$ & $\Delta N_{\rm eff}$& $\xi(T_{\rm BBN})$  \\   
           \hline
        (a) &1.50&0.692& 1.53E-5&0.0391 \\
        (b) & 1.36&0.675&1.22E-5&0.0369 \\
        (c) & 1.53&0.700&1.18E-2& 0.208\\
        (d) & 1.40&0.679 &6.37E-5&0.0558\\
        (e) & 1.43&0.684 &3.80E-4&0.0873\\
        (f) & 1.42&0.685&2.59E-5&0.0448\\
    \end{tabular}
    \caption{Table of $\Delta N_{\rm eff}$ and $\xi(T_{\rm BBN})=\left({T_h}/{T}\right)_{\rm BBN}$ when  $\xi_0=0.01$ and $\xi_0=1$ for 
    the model points of Table 1.
   As noted in the text the benchmarks of this table are 
    chosen to lie in the parameter space allowed in the 
    analysis of ref[44] which gives an exhaustive analysis of all
    of the current experimental constraints on the dark photon and its couplings and exhibits the parameter space still unconstrained.
    }
    \label{tab:my_label}
\end{table}

\subsection{Effect of $\xi_0$ on the allowed parameter space
and on spin-independent proton-DM cross section}
\begin{figure}[h]
\begin{center}
  \includegraphics[width=0.45\linewidth]{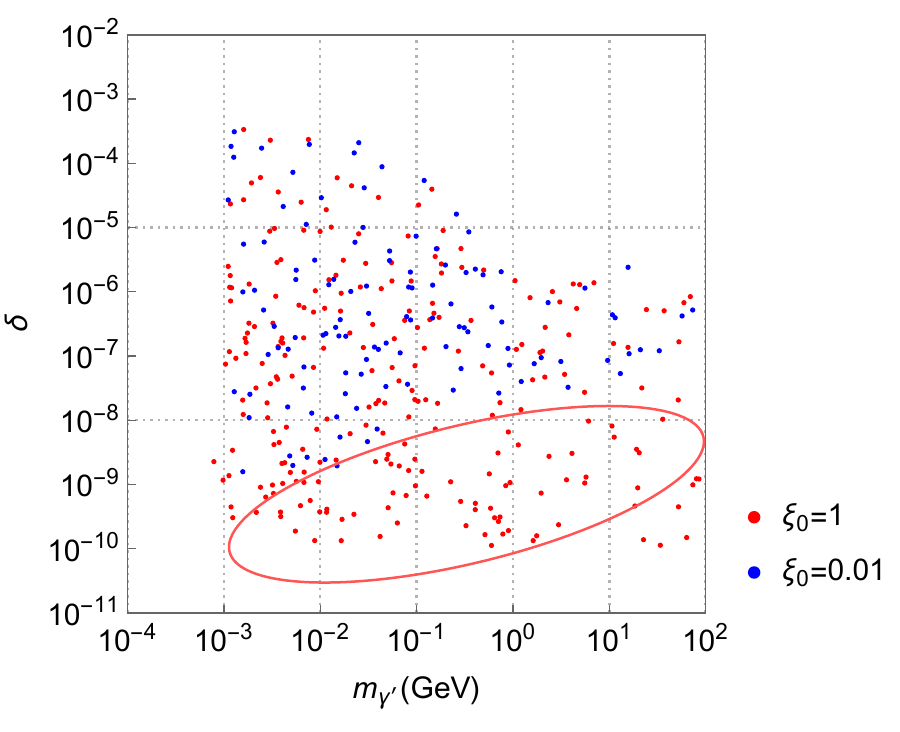}
     \includegraphics[width=0.35\linewidth]{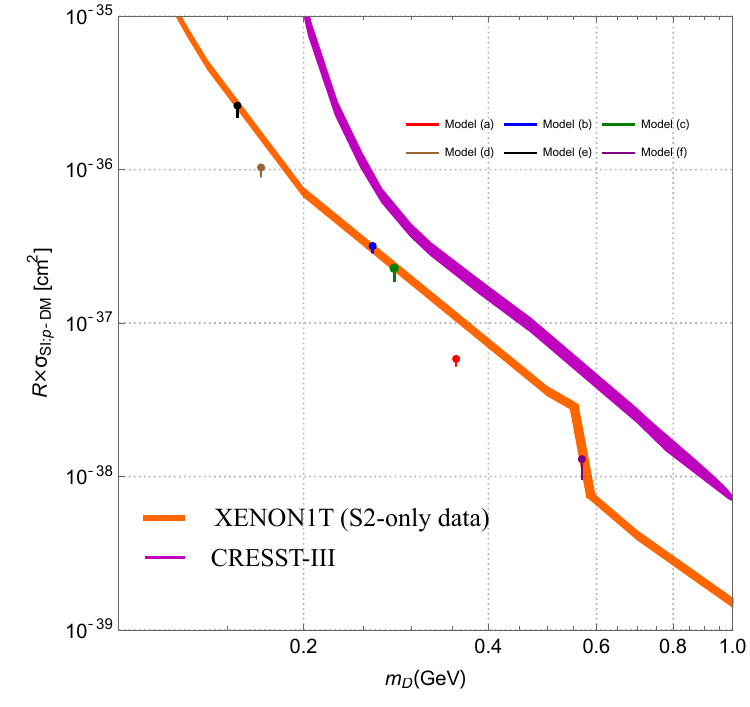}   
 \caption{ 
 Left panel:  A scatter plot of $\delta$ vs $m_{\gamma'}$  displaying the 
  models allowed 
under the constraint $0.012\leq \Omega h^2 \leq 0.12$
for  $\xi_0=0.01$ (blue) and $\xi_0=1$ (red).
The solid red ellipse shows that a significant region of the  parameter space in the $m_{\gamma'}-\delta$ plane   becomes accessible when $\xi_0=1$ 
 which would otherwise be excluded when $\xi=0.01$. This is meant as an illustration that $\xi_0$ plays a significant
 role in determining the allowed parameter space of models.
Right panel: Plot of the spin-independent proton-DM cross section 
for six model points where the vertical lines show the shift in the 
cross section as one moves from $\xi_0=0.01$ to $\xi_0=1$. 
The experiment constraints are from CRESST-III\cite{CRESST:2019jnq} and XENONIT\cite{XENON:2019zpr} }
\label{fig:ps5low}
\end{center}
\end{figure}

Next, we investigate the influence of $\xi_0$ on the allowed parameter space
consistent for a chosen range of relic density. To this end we constrain the relic
density to lie in the range $0.012 \leq \Omega h^2 \leq 0.12$ and 
 $m_D$ to lie in the range of  5 GeV to 10 TeV. Specifically we explore the allowed region for the two cases: $\xi_0=0.01$ and $\xi_0=1$. 
The result of our analysis is exhibited in the left panel of Fig.~\ref{fig:ps5low}
which gives a scatter plot
of the allowed models in $\delta$ vs $m_{\gamma'}$ where those with color blue 
correspond to $\xi_0=0.01$ and those with color red correspond to 
$\xi_0=1$.  
One of the interesting result
that emerges is that for $\xi_0=1$ most of the models lie in the range $10^{-10}<\delta<10^{-5}$ while 
for $\xi_0=0.01$ the allowed range is $10^{-9} <\delta <10^{-4}$. 
Thus, the analysis shows that the initial choice of $\xi_0$ significantly impacts the model's allowed parameter space.
$\xi_0$ also has significant effect on the  proton-DM  scattering
  cross section in the direct detection experiments for dark matter. 
 Specifically we  consider the spin independent proton-DM cross section
  $\sigma_{\rm SI: p-DM}$. 
Here we use the micrOMEGAs\cite{Belanger:2018ccd} to find the spin independent cross section.
  In the right panel of Fig.~\ref{fig:ps5low} we exhibit $\sigma_{\rm SI:p-DM}$ 
  for the six model points of Table \ref{tab:benchmarks} and their 
  dependence on $\xi_0$ in the range (0.01-1) is indicated by the 
  small vertical lines for each of the model points. 
  The numerical values of the  $\sigma_{\rm SI:p-DM}$  
  for $\xi_0=0.01$ and $\xi_0=1$ are exhibited in Table \ref{tab:SIcschange} 
  for  the models of Table \ref{tab:benchmarks}. Here one finds that
  the variation of the cross-section can be as large as 40\%. 
  Thus some of the models that are eliminated for the $\xi_0=1$ case
   would still be viable for the case $\xi_0=0.01$.   
      \begin{table}[h]
    \centering
    \begin{tabular}{lll}
        \hline
        Model & $\xi_0=1$ & $\xi_0=0.01$ \\
        & $\sigma_{\rm SI:p-DM}$(cm$^2$) & $\sigma_{\rm SI:p-DM}$(cm$^2$)  \\
           \hline
        (a) & 5.84E-38 &5.24E-38\\
        (b) & 3.18E-37& 2.87E-37 \\
        (c) & 2.19E-37 & 1.81E-37\\
        (d) &1.03E-36& 8.88E-37\\
        (e) & 2.61E-36& 2.23E-36\\
        (f) & 1.39E-38& 1.03E-38\\
    \end{tabular}
    \caption{Table of spin-independent proton-DM cross section
    $\sigma_{\rm SI:p-DM}$ for the model points of Table 1 for 
    $\xi_0=0.01$ and $\xi_0=1.0$.}
    \label{tab:SIcschange}
\end{table}
{
\section{Self-interacting dark matter, Sommerfeld enhancement,
and dependence on $\xi_0$}
\label{sec5}
\begin{figure}[h]
    \centering
   \begin{tikzpicture}
  \begin{feynman}
    \vertex at (0,0) (a) {\(\overline{D}\)};
    \vertex at (1,0) (b);
    \vertex at (2,0) (d) ;
    \vertex at (3,0) (f4) ;
    \vertex at (4,0) (f5){\(\overline{D}\)};
    
    \vertex at (1,-2) (c);
    \vertex at (0,-2) (f2) {\(D\)};
    \vertex at (2,-2) (f3) ;
    \vertex at (3,-2) (f6) ;
    \vertex at (4,-2) (f7)  {\(D\)};

    \diagram* {
      (f5) -- [fermion] (f4) -- [fermion] (d) -- [fermion] (b) -- [fermion] (a),
      (b) -- [boson, edge  label'=\(\gamma'\)] (c),
      (d) -- [boson, edge label'=\(\gamma'\)] (f3),
      (f4) -- [boson, edge label'=\(...\)] (f6),

      (c) -- [anti fermion] (f2),
      (c) -- [fermion] (f3) -- [fermion] (f6) -- [fermion] (f7) 
    };
  \end{feynman}
\end{tikzpicture}
    \caption{A diagram exhibiting a contribution to $D\bar D\to D\bar D$ scattering beyond the Born approximation.}
    \label{fig:schematicSE}
\end{figure}
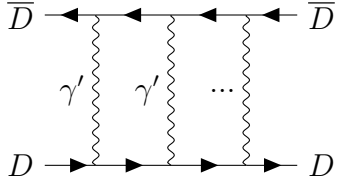
The self-interacting dark matter cross sections arises from the
processess $D\bar D\to D\bar D$, $DD\to DD$, and 
$\bar D \bar D\to \bar D \bar D$ via the exchange of a dark photon. 
The Lagrangian of Eq. (\ref{basic-lag}) leads to a Yukawa potential
between the D-fermions due to the dark photon exchange in the non-relativistic limit so that
\begin{align}
V(\vec{r})=\pm \frac{(g_X)^2}{4\pi }\frac{e^{-m_{\gamma'}r}}{r}\,,
\end{align}
where the plus sign is for $DD\to DD$ and $\bar D\bar D \to \bar D \bar D$
and the minus sign is for $D\bar D\to D\bar D$.
In some of the regions of parameters (i.e. $\frac{m_{\gamma'}}{m_D}\leq \frac{(g_X)^2}{4\pi }$), tree-level scattering or the  Born approximation is no longer valid and one has contributions from higher order dark photon 
exchanges as shown in  Fig. \ref{fig:schematicSE} which contribute to 
scattering. In this case we need to numerically solve the Schrodinger equation to find the accurate scattering cross sections. The radial equation one needs to solve is given by
\begin{align}
\left(\frac{d^2R_{l}}{dr^2}+\frac{2}{r}\frac{dR_{l}}{dr}-\frac{l(l+1)R_{l}}{r^2}\right)
+\left(p^2 -2\mu V(r)\right)R_{l}=0,
\label{Eq:SE}
\end{align}
where $p$ is the particle momentum and $V(r)$ is the potential.
The substitution $x = p r$ and  $R_{p,l} = Np\Phi_l(x)/x$ gives\cite{Iengo:2009ni}
\begin{align}
&\left(\frac{d^2}{dx^2}+1-\frac{l(l+1)}{x^2}- \frac{2ae^{-bx}}{x}\right)\Phi_l(x)=0,\non
&a=\pm \frac{\mu g_X^2}{4 \pi p}, ~~b= \frac{m_{\gamma'}}{p}.
\label{radial}
\end{align}
The  non-perturbative effect arising from the repeated  exchange of the 
mediator is often encoded in Sommerfeld enhancement and has been discussed
 in several previous works
  (see, e.g.,\cite{Lattanzi:2008qa,Arkani-Hamed:2008hhe,Cassel:2009wt,Cirelli:2007xd,Bringmann:2016din,Feng:2009mn} and the references therein). 
 Thus including non-perturbative effects the annihilation cross section times the velocity $v$ (where $v$ is the relative velocity in the CM system) for the  cross section $\sigma_{ab}$ for the elastic scattering process $a+b\to a+b$  may be  written as
 \beqn
 (\sigma_{ab} v)= S_E ({\sigma^0_{ab}} v)\,,
 \label{s.1}
 \eeqn
 where $({\sigma^0_{ab}} v)$ is the tree level cross section and 
 $S_E$ is the Sommerfeld enhancement.
 As noted in the present context the contribution 
 to the Sommerfeld enhancement arises from multiple exchanges
 of  the dark photon $\gamma'$. The solution of the differential equation Eq.(\ref{radial}) 
 has the form:
\begin{align}
    \Phi_l(x)_{x\rightarrow\infty} \rightarrow A\sin(x-\frac{l\pi}{2}+\delta_l),
    \label{eq:DEsol}
\end{align}
where $\delta_l$ is the phase shift for the $l-$th partial wave.
We write the Sommerfeld enhancement of $l$-th partial wave cross-section 
for the case of the Yukawa potential so that 
\begin{align}
    \sigma_{l}={S_{E}}_l\cdot\sigma_{0,l}\,,
\end{align}
where \cite{Iengo:2009ni},
    ${S_{E}}_l=({1\cdot 3\cdots(2l+1)}/{A})^2$.
Using Eq. (\ref{eq:DEsol}), we get
\begin{align}
A^2 &= A^2\sin^2(x-\frac{l\pi}{2}+\delta_l)+A^2\cos^2(x-\frac{l\pi}{2}+\delta_l)\non
 &= \Phi^2_l(x)_{x\rightarrow\infty}+ \Phi^2_l(x-\frac{\pi}{2})_{x\rightarrow\infty}\,,\non
{S_{E}}_l &= \frac{((2l+1)!!)^2}{\Phi^2_l(x)_{x\rightarrow\infty}+\Phi^2_l(x-\frac{\pi}{2})_{x\rightarrow\infty}}\,.
\end{align}
Taking $x$ larger than 30 gives a good enough approximation to the exact
solution. 
Typically an attractive potential
leads to  \se of cross section at low collision velocities, but 
one may also have Sommerfeld suppression for a repulsive potential.
\begin{figure}[h]
\begin{center}
  \includegraphics[width=0.4\linewidth]{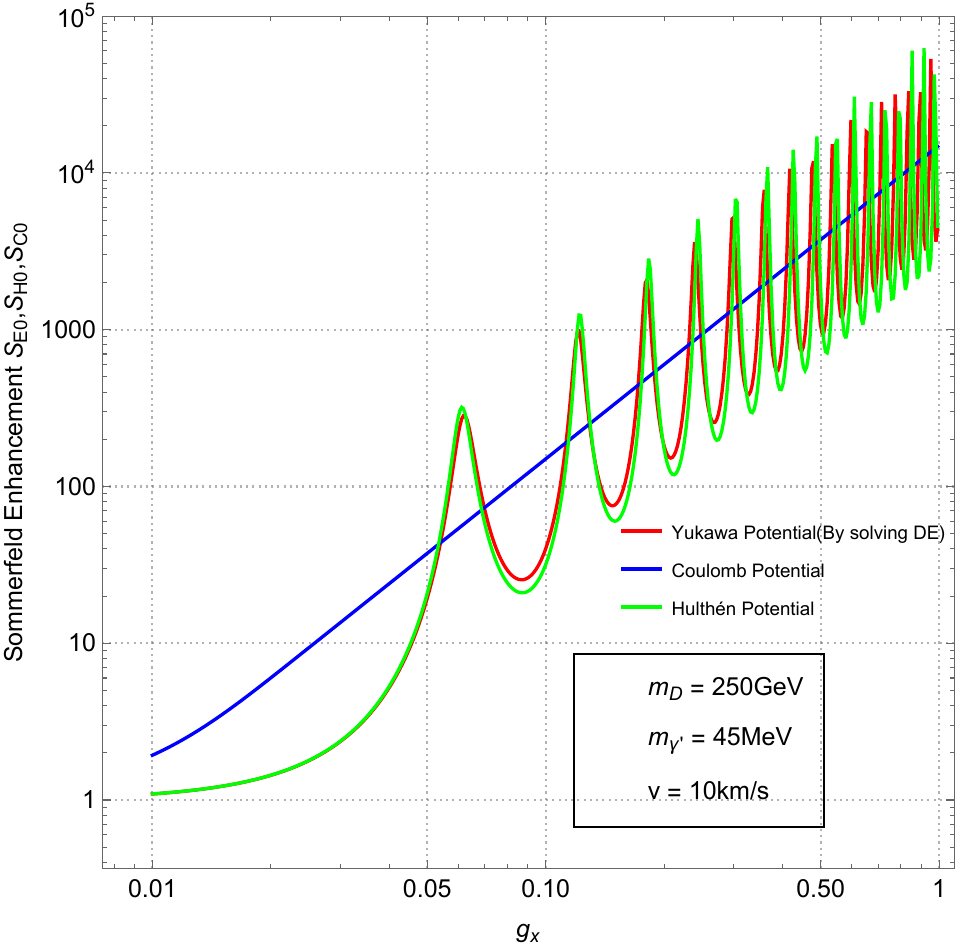} 
    \includegraphics[width=0.4\linewidth]{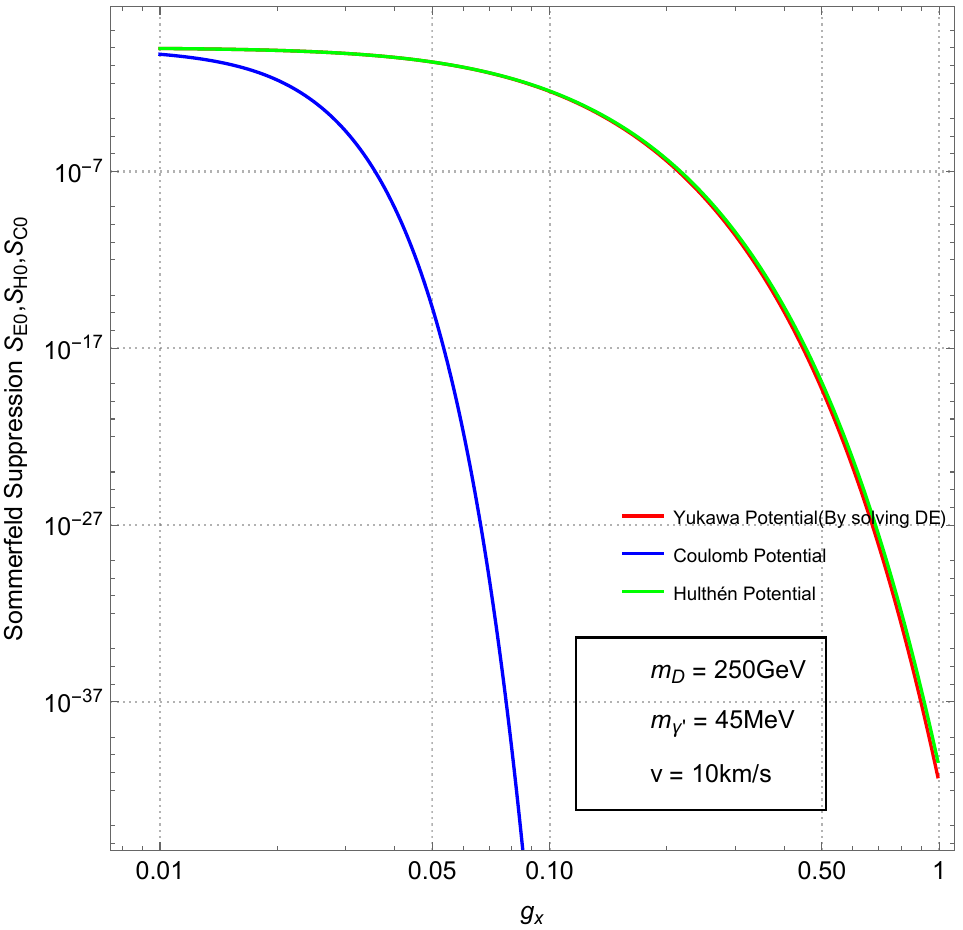}   
  \end{center}
\caption{
Left panel: Plot of S-wave Sommerfeld enhancement for an attractive
potential where  $S_{E0}$ is for the  Yukawa potential by solving numerically (red),
 $S_{C0}$ is for the Coulomb potential (black), and $S_{H0}$ is for the Hulthen 
 potential as a function of
$g_X$ for the case when $m_D=250$ GeV, $m_{\gamma'}=45$ MeV,
and $v=10$ km/s.
Right panel: Plot of S-wave Sommerfeld suppression for the case of 
a repulsive potential where  we use the same symbols
$S_{E0}, S_{H0}, S_{C0}$ for suppression as for enhancement to  avoid a proliferation of notation.}
\label{fig:hulthen}
\end{figure}
In the left panel of Fig. \ref{fig:hulthen} we exhibit \se for the case of a 
negative Yukawa potential.
Here we see that \se can be  very significant
and further the enhancement shows oscillatory behavior with $g_X$. 
To check the accuracy of our numerical
 analysis and to explain the oscillatory behavior 
  we compare our result with those from the Hulthen potential as an approximation to the
 Yukawa potential for which one can obtain a good analytic approximation 
 for the S-wave.
The Hulthen potential is given by\cite{Hulthen-1,Hulthen-2}
\begin{align}
    V({r}) = -\alpha \frac{\mu e^{-\mu r}}{1-e^{-\mu r}}\,, \quad  \mu = \frac{\pi^2 m_{\gamma'}}{6}\,, \quad\alpha = \frac{(g_X)^2}{4\pi }.
\end{align}
It is known that Hulthen potential is a very good approximation to Yukawa potential both at short and at long distances. With it one can find an analytic solution for 
the S-wave and thus find a good analytic approximation to the S-wave Sommerfeld enhancement \cite{Slatyer:2009vg,Kao:2020sqs}:
\begin{align}
    {S_{H0}} = \frac{\pi}{\epsilon_v}\frac{\sinh{\left(2\pi \epsilon_v
    \beta\right)}}{\cosh{\left(2\pi \epsilon_v\beta\right)}-
    \cos{\left(2\pi\sqrt{\beta-\epsilon_v^2\beta^2}\right)}}, \quad\epsilon_{v}=\frac{v}{2\alpha},\epsilon_{x}=\frac{m_{\gamma'}}{\alpha m_D},
    ~\beta=\frac{1}{\pi^2 \epsilon_x/6}\,.
    \label{Eq:Hulthen}
\end{align}
From  Eq.(\ref{Eq:Hulthen}), valid for the attractive potential case,
it is obvious that the oscillation is due to the existence of the cosine term. 
For the Coulomb potential the Sommerfeld enhancement for the S-wave is 
given by
\begin{align}
    S_{C0} = \pm\frac{2\pi\alpha}{v}\frac{1}{e^{\pm 2\pi \alpha/v} -1}
\end{align}
where plus is for repulsive potential and minus is for attractive potential.
The left panel of Fig.\ref{fig:hulthen}
gives a comparison of the S-wave Sommerfeld effect for three different potentials: Yukawa, Hulthen and Coulomb. The analysis shows that Hulthen potential gives a good approximation to the Yukawa potential and also explains the deep oscillations as a function of $g_X$. 
For the case of a repulsive potential ($\alpha$ negative),  the analysis is very different. A comparison of the numerical analysis using Yukawa potential 
and the analytic solution using Hulthen potential for the case of a repulsive
potential is given in the right panel of Fig. \ref{fig:hulthen}. Here again
one finds that the numerical analysis and the Hulthen potential result
 fully agree.
 
 Having checked the numerical accuracy of our analysis in Fig. \ref{fig:hulthen}
 we next investigate the effect of the Big Bang initial conditions on 
 Sommerfeld enhancement. In Fig. \ref{fig:Sommerfeldnegative} we compare S-wave \se for the cases
 $\xi_0=0.01$ and $\xi_0=1$ for the case of an attractive Yukawa potential. 
 The left panel of Fig. \ref{fig:Sommerfeldnegative} shows \se vs $v$ 
 and here one finds that $\xi_0=1$ (red) gives an enhancement which is
 larger than for the case $\xi_0=0.01$  (blue).  In the analysis we keep the relic density fixed at  $\sim 0.12$ for $\xi_0$ = 0.01 and $\xi_0=$1 by allowing
 $g_X$ to vary. 
 The right panel of 
  Fig. \ref{fig:Sommerfeldnegative} displays \se as a function of $m_{\gamma'}/m_D$ and here one finds that the oscillation peaks for the case $\xi_0=1$
  (red) are significantly larger than those for the case $\xi_0=0.01$ (blue). 
  A similar analysis for a repulsive Yukawa potential is carried out in  
  Fig. \ref{fig:Sommerfeldpositive}. However, in this case we have 
  Sommerfeld suppression rather than an enhancement where the
   Sommerfeld suppression is vs $v$ for the left panel and vs  
   $m_{\gamma'}/m_D$ for the right panel and  
    the red curve is for $\xi_0=1.0$ and the blue curve for $\xi_0=0.01$. 
    For both cases the Sommerfeld suppression is significantly larger 
  for  $\xi_0=1$ relative to  $\xi_0=0.01$. \\
  
\begin{figure}[h]
\begin{center}
  \includegraphics[width=0.4\linewidth]{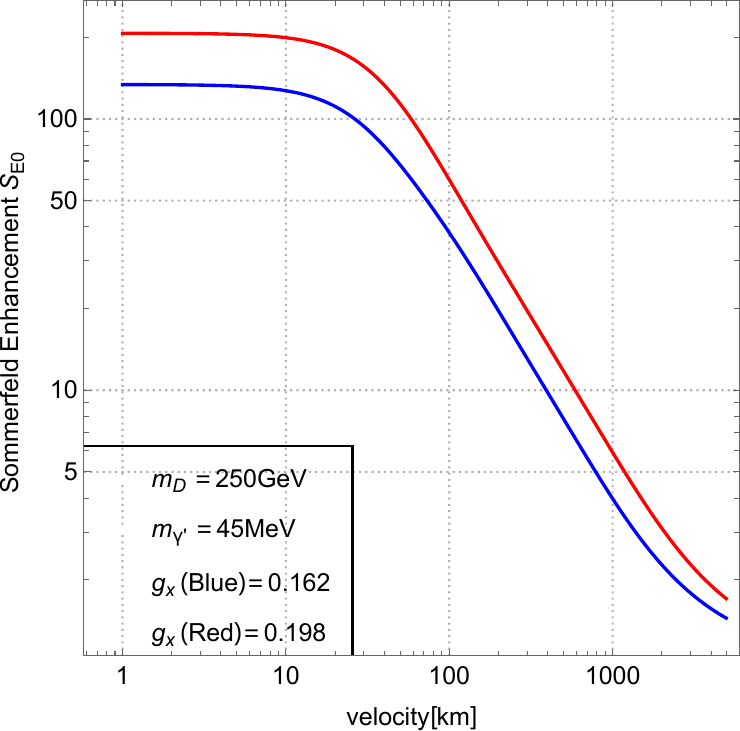} 
  \includegraphics[width=0.4\linewidth]{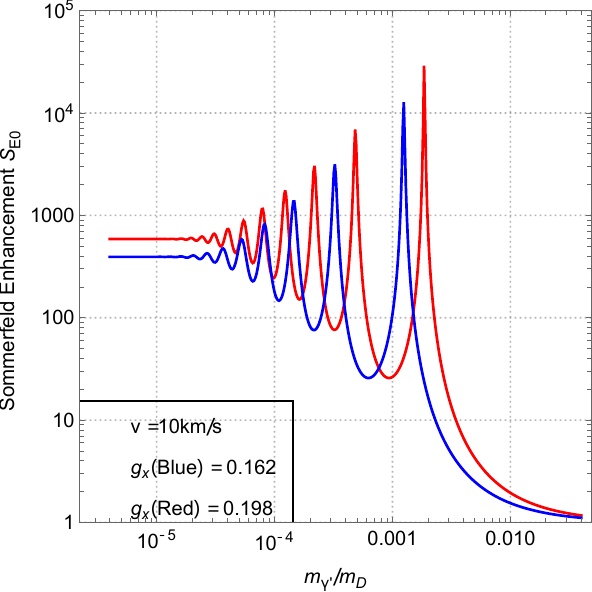} 
  \end{center}
\caption{
Plot of S-wave Sommerfeld enhancement for an attractive Yukawa potential for the case $\xi_0=0.01$ (blue) and for  $\xi_0=1$ (red) where for the left panel $x$-axis is $v$ and for the right panel the $x$-axis is  $m_{\gamma'}/m_D$. Here we allow $g_X$ to vary but keep the relic density $\sim 0.12$ for $\xi_0$ = 0.01 and $\xi_0=1$. 
}  
\label{fig:Sommerfeldnegative}
\end{figure}
\begin{figure}[h]
\centering
  \includegraphics[width=0.4\linewidth]{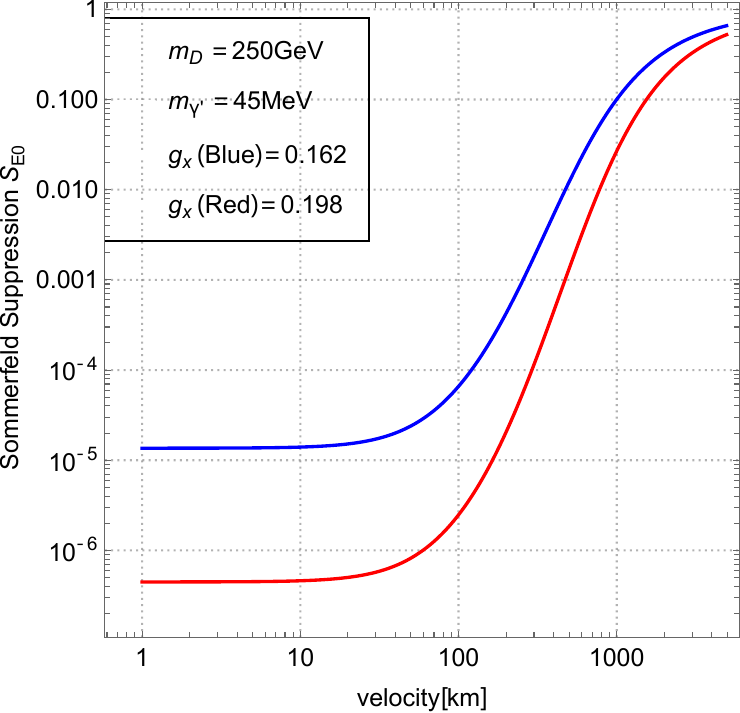}  
  \includegraphics[width=0.4\linewidth]{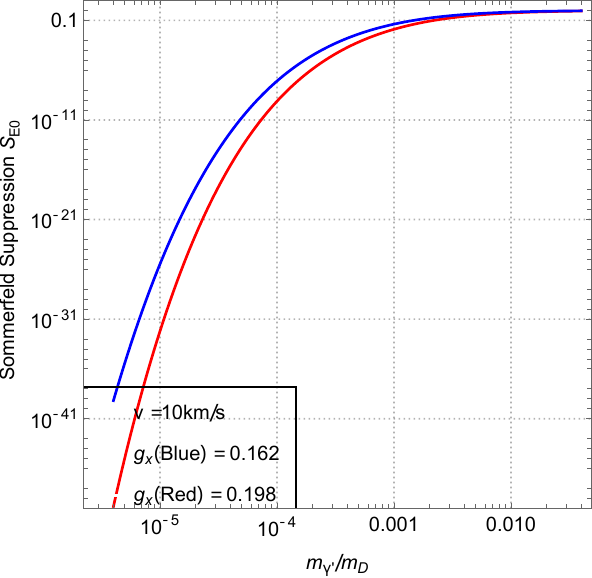} 
\caption{
Plot of S-wave Sommerfeld suppression for a repulsive Yukawa potential for the
case $\xi_0=0.01$ (blue) and for  $\xi_0=1$ (red) where for the left panel $x$-axis
is $v$ and for the right panel the $x$-axis is  $m_{\gamma'}/m_D$. Here we allow $g_X$ to vary to keep the relic density $\sim 0.12$ for $\xi_0$ = 0.01 and 
for $\xi_0=1$.}
\label{fig:Sommerfeldpositive}
\end{figure}
{
\section{Effect of $\xi_0$ on fit to galaxy data}
\label{sec6}
   Several analyses of galaxy data indicate that dark matter is collisional
   at the scale of dwarf galaxies and appears collision-less at the scale 
   of galaxy clusters\cite{Kaplinghat:2015aga,Sagunski:2020spe,Tulin:2017ara}.
   Thus, for dwarf galaxies one finds collisional velocity $\left<v\right>$ of dark
   matter in the range 10-100 km/s and 
   1  {cm}$^2$/g$<\sigma/m<50$ {cm}$^2$/g~\cite{Tulin:2017ara,Kaplinghat:2015aga} where $\sigma$ is the cross section and $m$ is the mass of DM particle.
   For midsize galaxies such as the low surface brightness galaxies (LSB) 
   and the Milky Way one finds $\left<v\right>$  in the range 80-200 km/s and 
   0.5  {cm}$^2$/g$<\sigma/m< 5 $ {cm}$^2$/g.  The galaxy clusters
      exhibit $\left<v\right> > 1000$ km/s. Here it is estimated that the $\sigma/m$ is maximally
      1 cm$^{2}$/g~\cite{Tulin:2017ara,Kaplinghat:2015aga,Robertson:2018anx}      
      and could be as low as 
   0.065  {cm}$^2$/g$<\sigma/m< 1 $ {cm}$^2$/g ~\cite{Elbert:2016dbb,Sagunski:2020spe,Andrade:2020lqq}.  
   As is well known one interesting possibility to account for the velocity dependence of the 
   DM cross sections is that DM is self-interacting  by Spergel and Steinhardt~\cite{Spergel:1999mh} and there is considerable follow up work on this idea
~\cite{Vogelsberger:2012ku,Rocha:2012jg,Peter:2012jh,Zavala:2012us,Elbert:2014bma,Vogelsberger:2014pda,Fry:2015rta,Dooley:2016ajo,Buckley:2009in,Loeb:2010gj,Tulin:2012wi,Tulin:2013teo,Schutz:2014nka,Bringmann:2016din}.   An analysis of 
    fit to the data within the dark photon model was previously done in 
   \cite{Aboubrahim:2020lnr} (see also \cite{Girmohanta:2022dog,Girmohanta:2022izb}).     
    Here we study the dependence
     of the fits on $\xi_0$. Further, here the analysis goes beyond the Born approximation
   used in  \cite{Aboubrahim:2020lnr} taking into account 
   non-perturbative effects encoded in the Sommerfeld enhancement 
   with also inclusion of identical particle exchange effects. 
   Since the dark matter is constituted of dark Dirac fermions 
   consisting of $D$ and $\bar D$ constituents, we will have processes
   of the type $DD\to DD$, $D\bar D\to D\bar D$ and $\bar D \bar D
   \to \bar D\bar D$.
   Thus the total cross section $\sigma_{DM}$ is given by
\begin{align}
\sigma_{DM}= \int d\Omega \left[ \frac{d\sigma_{D\bar D\to D\bar D}}{d\Omega}
+ \frac{1}{2} \frac{d\sigma_{DD\to DD}}{d\Omega} 
 +\frac{1}{2} \frac{d\sigma_{\bar D\bar D\to \bar D\bar D}}{d\Omega}
 \right]\,,
\end{align}
where the factor of $1/2$ arises due to identical nature of particles.

To numerically calculate the cross section, we start with Eq.(\ref{radial}) and 
use the method of \cite{Tulin:2013teo}.  
Here in the computation of DM cross sections, we need to calculate 
the phase shifts ($\delta_\ell$) for $D\bar D\to D\bar D$ separately from
the phase shifts ($\delta'_\ell$) for $DD\to DD$ while the phase
shifts for the process $\bar D \bar D\to \bar D \bar D$ will be the
same as for the process $DD\to DD$. Including all contributions, i.e.,
from $D\bar D\to D\bar D$,   $D D\to D D$ and  $\bar D\bar D\to D\bar \bar D$, and taking account of the identical 
nature of particles in $D D\to D D$ and  $\bar D\bar D\to D\bar \bar D$
scattering we find 
\begin{align}
 \sigma_{\text{tot}}=4\pi \sum_{\ell} (2\ell+1) \Big[|f_\ell|^2+ 2\left(1-\frac{1}{2} 
 (-1)^\ell \right)|f_\ell'|^2\Big]\,,
\label{phaseshifts}
\end{align}
where $f_\ell=e^{i\delta_\ell}\sin\delta_\ell/k$ and  $f'_\ell=e^{i\delta'_\ell}\sin\delta'_\ell/k$. 
The details leading to Eq.(\ref{phaseshifts}) are given  
 in section \ref{Appenix1}. The result of our numerical analysis to 
 fit the galaxy data on $\sigma v/m_D$  
 in the range of velocities from
 10 km/s to $10^4$ km/s is given in Fig. \ref{fig:galaxyfit}
  which exhibits the
  dependence of the fits on $\xi_0$ in the range $\xi_0=0.01$ to $\xi_0=1$. 
 In the analysis we allow $g_X$ to vary to keep the relic density 
 fixed at $\Omega_D h^2\sim 0.12$ 
  as  $\xi_0$ varies between 0.01 and 1.  The analysis shows that the variation of $\sigma v/m$ with $\xi_0$ is significant and  
  can sometimes   be as large as  $O(1)$ (see Model (f)) in Fig. \ref{fig:galaxyfit}. 
 We note that  the plots include Sommerfeld enhancement  
effects but these effects are relatively small. The reason for it is 
that the Sommerfeld enhancement  strongly depends on $g_X$ as can be
seen from the left panel of Fig. \ref{fig:hulthen}. However, in the analysis of galaxy 
fits of Fig. \ref{fig:galaxyfit}, we find that $g_X$ is relatively small which suppresses
the Sommerfeld enhancement. Our result here is consistent with a similar observation 
on Sommerfeld enhancement in the work of \cite{Bringmann:2016din} (see also \cite{Feng:2009mn}).
 In  Fig. \ref{fig:galaxyfit}
 the Born approximation results are also plotted for comparison
with the exact solutions. Further, we note that more fine tuned fits to the galaxy data can be gotten by adjustment of the model parameters such that resonances appear in some of the low lying partial waves, e.g., S, P and D waves. This is  exhibited in Fig. \ref{fig:ex4} where in the left panel  
we see enhancements in the S and the P waves appear to simulate the oscillations in the data at $\left<v\right>\sim 10^2$km/s and at $\left<v\right>\sim 10^3$km/s.
 On the right panel of Fig. \ref{fig:ex4}, we plot the cross section contributed from each partial wave separately. 
It is clear that the peak at $\left<v\right>\sim 10^2$km/s is largely due to
 the S-wave while the one at $\left<v\right>\sim 10^3$km/s  
 has a large contribution from $l=5$ although sum of all partial waves up 
 to $l=5$ enter in the fit given on the left panel.

\begin{figure}[h]
\captionsetup[subfigure]{labelformat=empty}
\begin{subfigure}{.3\textwidth}
  \centering
  \includegraphics[width=1\linewidth]{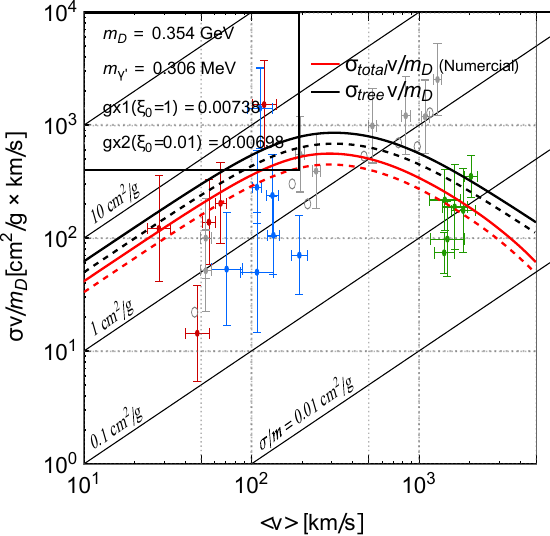}
  \caption{Model (a)}
  \label{fig:sfig1}
\end{subfigure}%
\begin{subfigure}{.3\textwidth}
  \centering
  \includegraphics[width=1\linewidth]{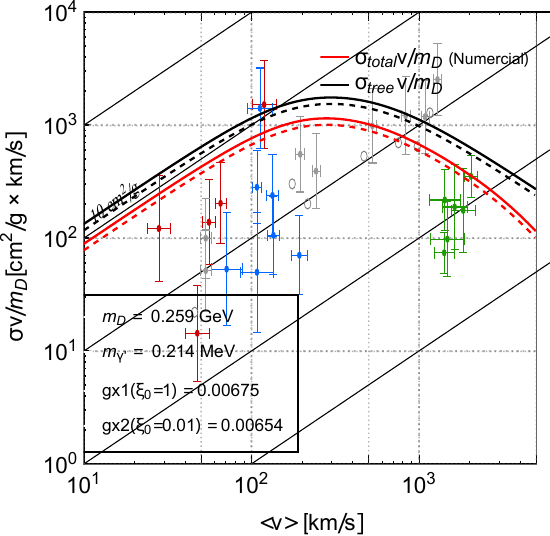}
  \caption{Model (b)}
  \label{fig:sfig2}
\end{subfigure}
\begin{subfigure}{.3\textwidth}
  \centering
  \includegraphics[width=\linewidth]{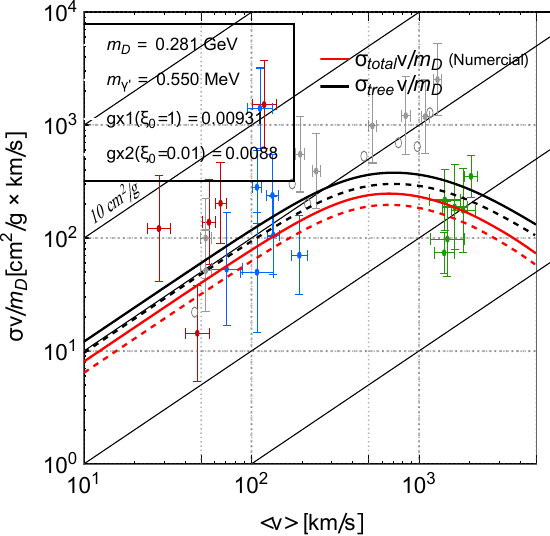}
  \caption{Model (c)}
  \label{fig:sfig2}
\end{subfigure}

\begin{subfigure}{.3\textwidth}
  \centering
  \includegraphics[width=1\linewidth]{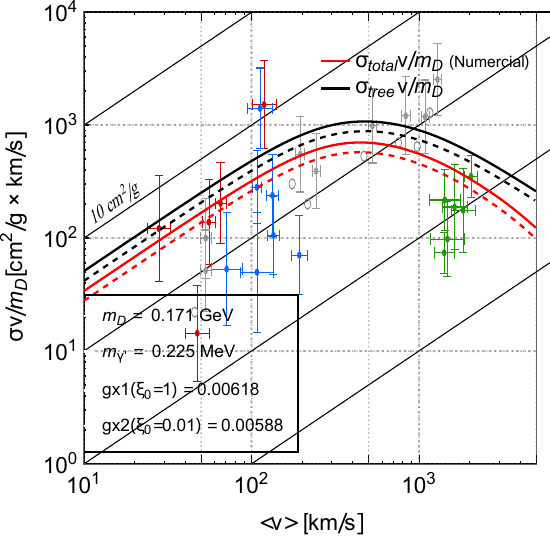}
  \caption{Model (d)}
  \label{fig:sfig2}
\end{subfigure}
\begin{subfigure}{.3\textwidth}
  \centering
  \includegraphics[width=1\linewidth]{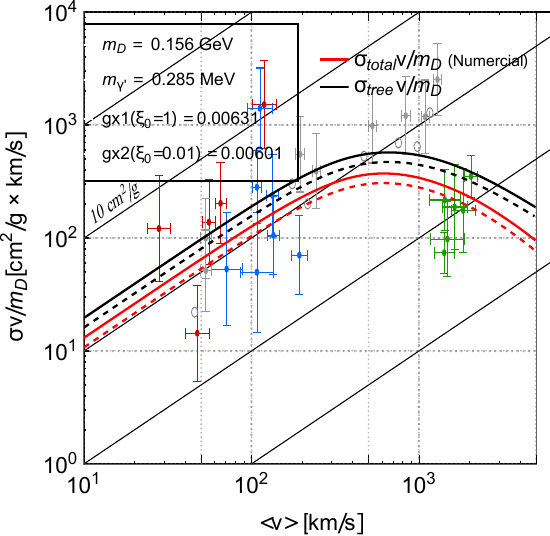}
  \caption{Model (e)}
  \label{fig:sfig2}
\end{subfigure}
\begin{subfigure}{.3\textwidth}
  \centering
  \includegraphics[width=1\linewidth]{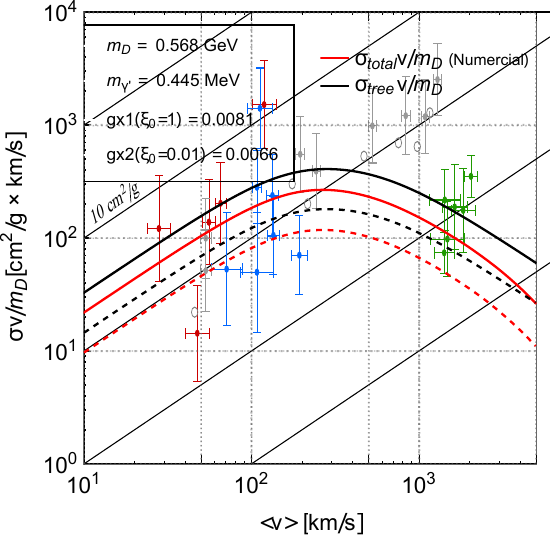}
  \caption{Model (f)}
\end{subfigure}
\caption{A fit to the galaxy data  
 taken from \cite{Kaplinghat:2015aga} which
studies the dependence of $\sigma v/m_D$ on $\xi_0$ in the range
    $(0.01-1)$ for the six models of Table 1. Here solid lines are for $\xi_0 =1$ and the dashed line for $\xi_0=0.01$ exhibiting the dependence of 
    $\sigma v/m_D$ 
   on $\xi_0$. The fits (in red)   are done using 
    the  full analysis by numerically integrating the Schrodinger equation
   including identical particle effects as well as Sommerfeld enhancement.
  For comparison we also exhibit the  tree-level QFT
   cross section shown by
  black curves which does not consider the effect of identical scattering. 
  In the analysis we allow $g_X$ to vary but keep the
  relic density  fixed at $\sim 0.12$ as 
  $\xi_0$ varies. It is to be noted that
``galaxy data'' is itself a computed quantity based on 
observation as evident from  \cite{Kaplinghat:2015aga}}  
   \label{fig:galaxyfit}
\end{figure}

\begin{figure}[h]
    \centering
    \includegraphics[width=0.4\linewidth]{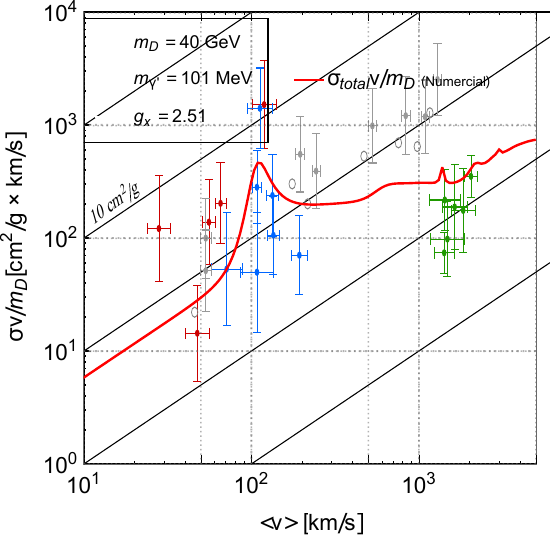}
    \includegraphics[width=0.4\linewidth]{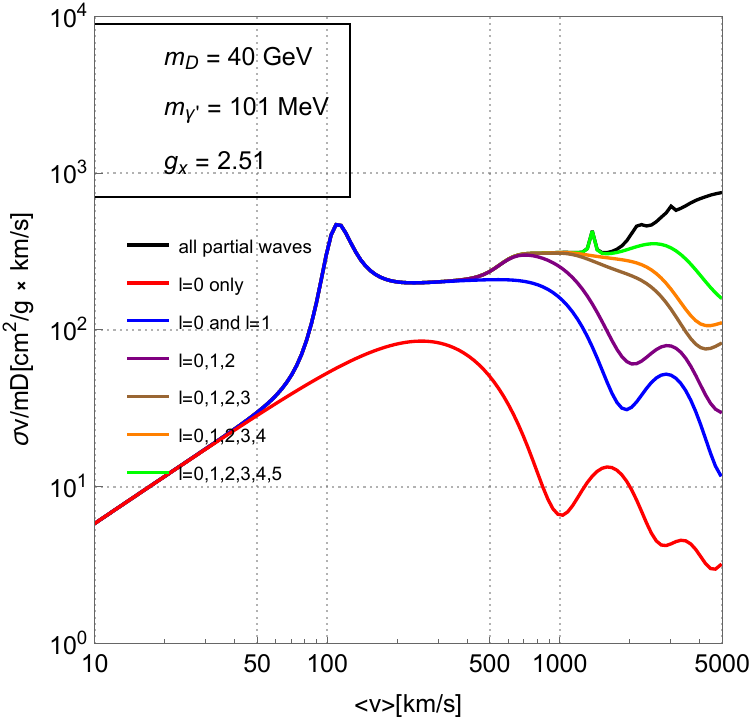}
    \caption{Left: Exhibition of peaks in fits to the galaxy data (which is the same as in Fig. 8) with 
    specific parameters. Right panel:  Here it is shown how the 
    peaks in the left panel at $\left<v\right>\sim 10^2$ km/s and at  $\left<v\right>\sim 10^3$ km/s arise from successive additional of higher waves. Thus the peak 
    at   $\left<v\right>\sim 10^2$ km/s arises mainly from S and P contributions, while the one at  $\left<v\right>\sim 10^3$ km/s    
     arises from contributions from up to $l=5$.}
    \label{fig:ex4}
\end{figure}
\newpage
Besides the total cross section, transfer cross section 
~\cite{Mohapatra:2001sx,Buckley:2009in,Ibe:2009mk,Loeb:2010gj,Tulin:2012wi,Tulin:2013teo,Schutz:2014nka}
($\sigma_{T}$) which have been used in the simulations of long range
interactions~\cite{Vogelsberger:2012ku,Zavala:2012us,Vogelsberger:2015gpr}.
Further, viscosity cross sections ($\sigma_{V}$) are also widely considered in 
analysis of SIDM\cite{Tulin:2013teo,Cline:2013pca,Boddy:2016bbu}.
They are defined so that 
\begin{align}
    \sigma_{total} =& \int \frac{d\sigma}{d\Omega} d\Omega\,, \non
    \sigma_{T} =& \int \frac{d\sigma}{d\Omega} (1-\cos{\theta)} d\Omega\,, \non
    \sigma_{V} =& \int \frac{d\sigma}{d\Omega} (1-\cos^2{\theta)} d\Omega\,.
    \label{totTV}
  \end{align}
In terms of partial waves $\sigma_T$ and $\sigma_V$ are given by
  \begin{align}
  \sigma_{\text{T}}=&4\pi \sum_{\ell} \Big[T(f_\ell)+ 2\left(1-\frac{1}{2} 
 (-1)^\ell \right)T(f_\ell')\Big]\label{cross2}\,.\\
  \sigma_{\text{V}}=&4\pi \sum_{\ell} \Big[V(f_\ell)+ 2\left(1-\frac{1}{2} 
 (-1)^\ell \right)V(f_\ell')\Big]\,.\\
 T(f_\ell)=&\left((2l+1)|f_l|^2- l f_l f^{*}_{l-1}-(l+1) f_l f^{*}_{l+1}\right)\,.\\
 V(f_\ell)=&\left(\frac{2(l^2+l-1)(2l+1)}{(2l-1)(2l+3)}|f_l|^2- \frac{(l-1)l}{(2l-1)}f_l f^{*}_{l-2}- \frac{(l+2)(l+1)}{(2l+3)}f_l f^{*}_{l+2}\right)\,.
\label{cross3}
\end{align}
Details of their computation in terms of partial waves are given in section \ref{Appenix1}. Fig.~\ref{fig:xianalysis3}
shows that 
 $\sigma_{tot}, \sigma_T,
\sigma_V$ differ significantly from each other. We also exhibit the tree level cross section
for comparison.
\begin{figure}
    \centering
    \includegraphics[width=0.5\linewidth]{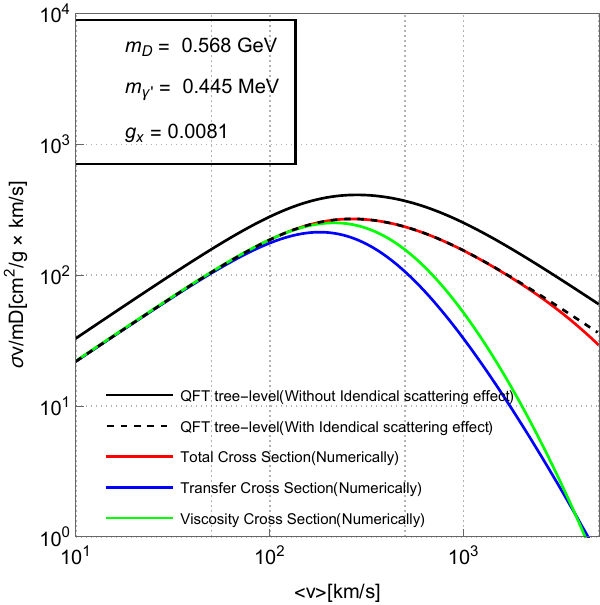}
    \caption{A comparison of  $\sigma v/m_D$ vs $v$ for different cross sections which include the total cross section
    $\sigma_{total}$, the transverse cross section  
    $\sigma_T$, the viscosity cross section $\sigma_V$ as defined by 
    Eq. (\ref{totTV}) and the tree level QFT cross section with and without
    identical particle effects for model (f) of Table  \ref{tab:benchmarks}. The analysis is done for $\xi_0=1$. }
   \label{fig:xianalysis3}  
\end{figure}
\section{Conclusion}\label{sec7}
Hidden sectors are ubiquitous in models of extra dimensions, in 
extended supergravity and in strings and appear 
 in a variety of  beyond the standard model constructions
such  as in moose/quiver gauge theories (see, e.g.,~\cite{Rothstein:2001tu,Douglas:1996sw,Arkani-Hamed:2001kyx,Hill:2000mu}).  
 While the hidden sectors are neutral under the standard model
 gauge group they can couple feebly with the standard model.
 However, the couplings of the hidden sector to the inflaton 
could vary over a wide range. Thus on one extreme the hidden sector
coupling to the inflaton could be negligible relative to the coupling
of the standard model. In this case at the end of reheating 
there would be essentially no production of the hidden sector particles, except
via gravitational production,
and the hidden sector would likely be colder than the standard model.
On the other extreme the hidden sector and the visible sectors could
couple democratically to the inflaton and in this case the hidden 
sector and the visible sectors would be in thermal equilibrium at the
end of reheating. These two extremes would have significantly different
thermal evolution and would result in significant differences in their
predictions of the physical observables. In this work we have investigated
these effects in the context of a specific hidden sector model which
arises from a $U(1)_X$ extension of the standard model gauge group.
The  contents of  the hidden sector consists of a 
 dark fermion
which has gauge interactions with the $U(1)_X$ gauge field. The 
communication between the hidden sector and the visible sector 
arises from kinetic mixing between the $U(1)_X$ and $U(1)_Y$ gauge
fields where the $U(1)_X$ gauge field acquires mass via the Stueckelberg
mechanism. In view of the asymmetric coupling of the visible and the hidden sectors to the inflaton field, the temperature of 
the hidden sector $T^0_h$ and of the visible sector $T^0$ will in general 
be unequal at the end of reheating. Thus the ratio of the two, i.e., $\xi_0=(T^0_h/T^0)$ enters in the thermal evolution of the hidden and the visible sectors and affects phenomena at low energy.  

The analysis of the work provides a cosmologically consistent 
framework in that it involves a synchronous evolution of the coupled  hidden and visible sectors.
  In the above framework we investigate a number of phenomena and their dependence on $\xi_0$. These include dark freeze-out, relic density 
and the extra number of relativistic degrees of freedom at the
BBN time, and the proton-DM cross-section. Further, we investigate
the effects of $\xi_0$ on the self-interaction
cross section and on Sommerfeld enhancement. 
The model is then used in fitting self-interacting dark matter cross sections from galaxy scales to the scale of galaxy clusters. Here we find that fits to data show a significant variation sometime as much as $O(1)$ for $\xi_0$ in the range $(0,1)$. Thus the analysis indicates that inclusion of hidden sectors 
which appear in a variety of models of particle physics beyond the standard model, the initial constraints on the hidden sector at the end of reheating and specifically on $\xi_0$ could have significant influence on observables and thus their inclusion will be relevant for accurate description of physical phenomena.
While our analysis is done for the case of one portal, the general techniques
discussed here would be valid for a broader class of models.
Finally, we show that the approximation often made in 
 the thermal evolution of visible and hidden sectors by assuming 
  entropy conservation for each of the sectors separately
 gives widely inaccurate results even for the case for very feeble 
 interactions such as, for example, with the kinetic mixing parameter as low as $\delta =10^{-10}$. Such an analysis is  thus a poor approximation to the analysis we carry out for the thermal evolution of the
 visible and the hidden sectors in a synchronous manner using
 Eq.(\ref{DE1}). For generality we also consider the case with the mass mixing parameter $\epsilon$ which reaches a similar conclusion.
 We have also analyzed the accuracy of assuming the conservation of total entropy
 for the yield equations and find that the differences between conservation assumption
 and no conservation assumption are typically within $O(15\%)$. We conclude that
  an accurate thermal evolution is essential for the current and future precision analyses in cosmology while analyzing physics involving hidden sectors.
 \\

\noindent
{\bf Acknowledgments:}  Discussions with Amin Abou Ibrahim and Zhuyao
Wang are acknowledged. The research of JL and PN was supported in part by the NSF Grant PHY-2209903.

\section{Appendix A: Source functions \label{appen:A}}
The source term $j_h$ that appears in Eq.(\ref{DE1})
 is defined by:
\begin{align}
    j_h = &\sum_i\Big[2Y_i^{eq}(T)^2J(i\bar{i}\rightarrow D\bar{D})(T)
    +Y_i^{eq}(T)^2J(i\bar{i}\rightarrow \gamma')(T)\Big]\s^2
    -Y_{\gamma'}J(\gamma'\rightarrow e^+e^-)(T_h)\s\,.
    \label{eq:sourceterm}
\end{align}
The $J$-functions that appear in Eq.~(\ref{eq:sourceterm}) are defined as 
\begin{align}
&n^{\rm eq}_i(T)^2 J(i~\bar{i}\to D\bar{D})(T)
=\frac{T}{32\pi^4}\int_{s_0}^{\infty}ds~\sigma_{D\bar{D}\to i\bar{i}}s(s-s_0)K_2(\sqrt{s}/T)\,, \\
&n^{\rm eq}_i(T)^2 J(i~\bar{i}\to \gamma')(T)
=\frac{T}{32\pi^4}\int_{s_0}^{\infty}ds~\sigma_{i\bar{i}\to \gamma'}s(s-s_0)K_2(\sqrt{s}/T)\,. 
\end{align}
\begin{equation}
n_{\gamma'}J(\gamma'\to e^+ e^-)(T_h)=n_{\gamma'}m_{\gamma'}\Gamma_{\gamma'\to e^+ e^-}\,,
\end{equation}
and
\begin{align}
&n_i^{\rm eq}(T)^2\langle\sigma v\rangle_{i\bar{i}\to\gamma'}(T)
= \frac{T}{32\pi^4}\int_{s_0}^{\infty} ds ~\sigma(s) \sqrt{s}\, (s-s_0)K_1(\sqrt{s}/T)\,,
\end{align}
where as noted earlier $K_1$ is the modified Bessel function of the second kind and degree one and $s_0$ is the minimum value of the Mandelstam variable $s$.
{{
\section{Appendix B: Model details \label{appen:Modeldetial}\label{append:B}}
 In addition to the interactions given in section \ref{sec3}, there are  interactions
  involving the dark sector and the standard model particles in the canonical basis 
  where the kinetic energy and the mass matrices of the gauge boson are diagonal.
   Here the standard model fermions (i.e., quarks and leptons) have 
 feeble interactions with the dark photon which are given by 
\begin{align}
\label{D-darkphoton}
\Delta \mathcal{L}_{\rm int}&=
 \frac{g_2}{2\cos\theta}\bar\psi_f\gamma^{\mu}\Big[(v'_f-\gamma_5 a'_f)A^{\gamma'}_{\mu}\Big]\psi_f.  
\end{align}
where $g_2$ is the $SU(2)_L$ gauge coupling constant,
 $f$ stands for the standard model fermions and angle $\theta$
is defined in Eq.(\ref{angles}).
The  vector and axial vector couplings of the dark photon 
with the SM fermions $f$ are given by
\begin{equation}
\begin{aligned}
v'_f&=-\cos\psi[(\tan\psi-s_\delta\sin\theta)T_{3f}-2\sin^2\theta(-s_{\delta} \csc\theta+\tan\psi)Q_f],\\
a'_f&=-\cos\psi(\tan\psi-s_{\delta} \sin\theta)T_{3f}.
\end{aligned}
\label{eqn:v-a}
\end{equation}
Here $s_\delta= \sinh\delta$, $T_{3f}$ is the third component of isospin, and $Q_f$ is the electric charge for the fermion $f$.  
The angles $\theta$ and $\psi$ which along with $\phi$ are the three
Euler angles with diagonalize the $3\times 3$ gauge boson mass square
matrix involving the fields $C^\mu, B^\mu, A^\mu_3$ are defined as
\cite{Feldman:2007wj}
\begin{align}
    \tan{\phi} = - s_{\delta},\quad \tan{\theta} = \frac{g_Y}{g_2}c_{\delta}\cos{\phi},\quad \tan{2\psi} = \frac{-2s_{\delta} m_Z^2\sin{\theta}}{m_{\gamma'}^2-m_Z^2+(m_{\gamma'}^2+m_Z^2-m_W^2)\delta^2}\,,
    \label{angles}
\end{align}
where $c_{\delta}=\cosh{\delta}$.  
For the model of 
Eq.(\ref{basic-lag}) where the hidden sector consists of the dark
photon and a dark Dirac fermion, $g^h_{\rm eff}$ and  $h^h_{\rm eff}$ 
are given by
 {
\begin{align}
g^h_{\rm eff} &= g^{\gamma'}_{\rm eff}+g^D_{\rm eff}\,, && h^h_{\rm eff}=h^{\gamma'}_{\rm eff}+h^D_{\rm eff}, \nonumber\\
g^{\gamma'}_{\rm eff}&=\frac{45}{\pi^4}\int^{\infty}_{x_\gamma'}\frac{\sqrt{x^2-x_{\gamma'}^2}}{e^x-1}x^2dx, && 
h^{\gamma'}_{\rm eff}=\frac{45}{4\pi^4}\int^{\infty}_{x_\gamma'}\frac{\sqrt{x^2-x_{\gamma'}^2}}{e^x-1}(4x^2-x_{\gamma'}^2)dx,\nonumber\\
g^{D}_{\rm eff}&=\frac{60}{\pi^4}\int^{\infty}_{x_D}\frac{\sqrt{x^2-x_{D}^2}}{e^x+1}x^2dx, &&  h^{D}_{\rm eff}=\frac{15}{\pi^4}\int^{\infty}_{x_D}\frac{\sqrt{x^2-x_{D}^2}}{e^x+1}(4x^2-x_{D}^2)dx\,,
\label{geff-heff}
\end{align}
}
 where $x_D= m_D/T_h$, $x_{\gamma'}= m_{\gamma'}/T_h$. 

Further, to compute $\zeta_h =\frac{3}{4}(1+\frac{p_h}{\rho_h})$, we need  $\rho_h$ and $p_h$, which are given by
\begin{align}
\rho_h &= \rho_{\gamma'} +\rho_D , && p_h = p_{\gamma'}+ p_D, \nonumber\\
\rho_{\gamma'}&=\frac{g_{\gamma'}T^4}{2\pi^2}\int^{\infty}_{x_\gamma'}\frac{\sqrt{x^2-x_{\gamma'}^2}}{e^x-1}x^2dx, && 
p_{\gamma'}=\frac{g_{\gamma'}T^4}{6\pi^2}\int^{\infty}_{x_{\gamma'}}\frac{(x^2-x_{\gamma'}^2)^{\frac{3}{2}}}{e^x-1}dx,\nonumber\\
\rho_{D}&=\frac{g_{D}T^4}{2\pi^2}\int^{\infty}_{x_D}\frac{\sqrt{x^2-x_{D}^2}}{e^x+1}x^2dx, && 
p_{D}=\frac{g_{D}T^4}{6\pi^2}\int^{\infty}_{x_D}\frac{(x^2-x_{D}^2)^{\frac{3}{2}}}{e^x+1}dx.
\label{rhoh-ph}
\end{align}
Here $g_{\gamma'} = 3$ and $g_D = 4$.
For total $\zeta$,  we need energy and pressure densities for 
both the visible and the hidden sectors and we use the relation
\begin{align}
    \zeta = \frac{3}{4}(1+\frac{p_v+p_h}{\rho_v+\rho_h}).
\end{align} 
Before proceeding further we note that the extension to include
both the kinetic and mass mixings in Eqs.(\ref{ddgammap})-(\ref{angles}) is straightforward as has been discussed in \cite{Feldman:2007wj}
and we exhibit them here for easy reference to guide the 
discussion in section \ref{sec7}.
With inclusion of both the kinetic mixing parameter
$\delta$ and the mass mixing parameter $\epsilon=M_2/M_1$, the neutral 
current Lagrangian in the hidden sector takes the form

\begin{align}
\mathcal{L}_{NC}^{\rm{hid}} &= \bar{D}\gamma^{\mu}
                        \left[ \epsilon_{\gamma'}^{D} A_{\mu}^{\gamma'}+\epsilon_{Z}^{D} Z_{\mu}+\epsilon_{\gamma}^{D}
                        A^{\gamma}_{\mu}\right]D.
   \label{delta-epsilon-1}
\end{align}
where 
\begin{align}
    \epsilon_{\gamma'}^{D}        &
                                   \simeq      g_X Q_X,~~
    \epsilon_{Z}^{D}    
                                    \simeq     \bar\epsilon g_X Q_X \sin \theta  \left[  1+\frac{\delta}{\bar\epsilon} \right],~~
    \epsilon_{\gamma}^{D}    
                                    \simeq     -\bar\epsilon g_X Q_X \cos \theta  \left[  1+\frac{\delta}{\bar\epsilon} \right].
  \label{delta-epsilon-2}
\end{align}  
where $\bar \epsilon$ is defined so that $\bar \epsilon=(\epsilon-\delta)/\sqrt{1-\delta^2}$. There is also a corresponding modification
of the neutral current in the visible sector as discussed in \cite{Feldman:2007wj}. The constraints on $\delta$ and $\epsilon$
arising from the fits to the electroweak data are mild and one
finds that $|\epsilon-\delta|$ can be as large as 0.05 consistent with
the same level of $\chi^2$ fits to the electroweak data as the standard model\cite{Feldman:2007wj}.
}}
\section{Self-interacting dark matter cross sections
\label{appen:C}} 

\label{Appenix1}

\subsection{$D\bar D\to D\bar D$}
Let us first consider $D\bar D\to D \bar D$ scattering. Here the wave function
for scattering of  a plane wave scattering from a central potential is given 
by $\psi(\vec r) \sim  [e^{ikz} + f(\theta)e^{ikr}/{r}]$.
In this case the scattering amplitude $f(\theta)$ has  an 
expansion in terms of the partial wave amplitudes 
$f_l=\frac{e^{i \delta_l} \sin\delta_l}{k}$ where $\delta_l$ is the phase
shift for the l-th partial wave so that $f(\theta)$ has the expansion
$f(\theta) = \sum_{l=0}^{\infty}  (2l+1) f_l P_l(\cos\theta)$
and $\sigma_{\text{tot}}^{DD}$
have the expansion
\begin{align}
 \sigma^{D \bar{D}}_{total}&=  \int  |f(\theta)|^2 d\Omega=(4\pi)   \sum_{l} (2l+1) |f_l|^2\,.
\end{align}
The transfer cross-section is defined by
\begin{align}
 \sigma^{D \bar{D}}_T&=  \int  |f(\theta)|^2 (1-\cos{\theta)} d\Omega\,. 
 \end{align}
Using the relations 
\begin{align}
\int (dx)x P_{l} (x) P_{l'}(x)&= \left\{\begin{array}{c}
    \frac{2(l'+1)}{(2l'+1)(2l'+3)}\quad \textrm{for }l=l'+1  \\
 \frac{2l'}{(2l'-1)(2l'+1)}\quad \textrm{for }l=l'-1\,,     
\end{array} \right.\
\end{align}
 the transfer cross section can be written as 
\begin{align}
 \sigma^{D \bar{D}}_T&= 4\pi\sum_{l=0}^{\infty}\left((2l+1)|f_l|^2- l f_l f^{*}_{l-1}-(l+1) f_l f^{*}_{l+1}\right)\,.
\end{align}

The viscosity cross-section defined by
\begin{align}
 \sigma^{D \bar{D}}_V&=  \int  |f(\theta)|^2 (1-\cos^2{\theta)} d\Omega\,,
\end{align}
can be expanded in terms of partial waves using the relations 
\begin{align}
\int (dx)x^2 P_{l} (x) P_{l'}(x)&=\left\{\begin{array}{c}
    \frac{2(l'+1)(l'+2)}{(2l'+1)(2l'+3)(2l'+5)}\quad \textrm{for }l=l'+2  \\
 \frac{2(2l'^2+2l'-1)}{(2l'-1)(2l'+1)(2l'+3)}\quad \textrm{for }l=l'\\
 \frac{2l'(l'-1)}{(2l'-1)(2l'+1)(2l'-3)}\quad \textrm{for }l=l'-2\,,\\ 
\end{array} \right.\
\end{align}
which gives 

\begin{align}
 \sigma^{D \bar{D}}_V
 &= 4\pi\sum_{l=0}^{\infty}\left(\frac{2(l^2+l-1)(2l+1)}{(2l-1)(2l+3)}|f_l|^2- \frac{(l-1)l}{(2l-1)}f_l f^{*}_{l-2}- \frac{(l+2)(l+1)}{(2l+3)}f_l f^{*}_{l+2}\right)\,.
\end{align}

\subsection{$D D\to D D$, $\bar D\bar D\to \bar D\bar D$}          
Here the scattering involves identical particles which are fermions
so the overall wave function for the particles must be anti-symmetric.
This can happen in two ways: (i) spin anti-symmetric and space symmetric
or (ii) spin symmetric and space anti-symmetric. Now the two spin particles
can have a total spin 1 (triplet state) or total spin is zero (singlet state).
For the triplet state the space wave function must be anti-symmetric in 
$\theta\to \pi-\theta$ and for the singlet state the space wave function must be symmetric.  Thus we have for $\sigma^{DD}(\theta)$ 
the expression 

\begin{align}
\sigma^{DD}(\theta)&= \frac{3}{4}|f'(\theta)-f'(\pi-\theta)|^2 + \frac{1}{4}  |f'(\theta)+f'(\pi+\theta)|^2 \non
&=  |f'(\theta)|^2+ |f'(\pi-\theta)|^2 
- Re(f'(\theta) f^{'*}(\pi-\theta))\,.
\label{D.1}
\end{align}
Further, $\sigma_{\text{tot}}^{DD}$ is given by
\begin{align}
\sigma_{\text{tot}}^{DD}= \frac{1}{2} \int \sigma^{DD}(\theta) d\Omega\,,
\end{align}
where the front factor of $1/2$ is to take account of the identical nature
of the scattering particles.
The partial wave analysis of $\sigma_{\text{tot}}^{DD}$ gives
\begin{align}
\sigma_{\text{tot}}^{DD}
&= 2\pi \sum_l(2l+1) \left[|f_l|^2     + 2\left [ 1 -\frac{1}{2}(-1)^l\right]  |f'_l|^2  \right]\,.
\label{2.} 
\end{align}
Here $f_l'= e^{i\delta'_l} \sin\delta'_l$ since the potential governing 
$DD\to DD$ scattering is different from the one that governs
$D\bar D\to D\bar D$ scattering.
Similar calculations are done for transfer cross section and 
for viscosity cross section. Further, 
$\sigma_{\text{tot}}^{\bar D\bar D}=\sigma_{\text{tot}}^{DD}$.

\section{Entropy conservation approximation\label{appen:D}}
\label{sec:entropy} 
Here in subsection \ref{ent1} we will discuss the validity of separate entropy conservation approximation for visible and hidden sectors and in subsection \ref{ent2} we will
discuss the validity of the conservation of the total entropy which is the sum of the visible 
and the hidden sector entropies. 

\subsection{On the validity of separate entropy conservation approximation 
of visible and hidden sector\label{ent1}}

 In several previous works (see, e.g.,\cite{Feng:2008mu})
  an assumption of entropy conservation per co-moving volume
 separately for the visible and the hidden sectors is made to relate $\xi(T)$ at different temperatures. The above implies that the ratio $s_h/s_v$ is unchanged at different temperatures where $s_v$ and $s_h$ are the entropy densities for the visible and the hidden sectors where 
\begin{align}
s_v= \frac{2\pi^2}{45} h^v_{eff} T^3, ~s_h=\frac{2\pi^2}{45} h^h_{eff} T_h^3.
\end{align}
Specifically it is assumed that the following relation between the temperatures $T_0$ and $T$ holds
\begin{align} 
\frac{h^h_{eff}(T_h)}{h^v_{eff}(T)}\xi^3(T)
=\frac{h^h_{eff}(T_{0h})}{h^v_{eff}(T_0)}\xi^3(T_0)
\label{entropy-con}
\end{align}
Noting that $T_{h}=\xi(T) T$ and $T_{0h}= \xi_0 T_0$ where 
$\xi_0\equiv \xi(T_0)$ we can write the above equation as follows 
\begin{align}
\left(h_{eff}^h(\xi(T) T)\right)^{1/3}\xi(T)
= (h^v_{eff}(T))^{1/3} 
\left(\frac{h^h_{eff}(\xi_0 T_0)}{ h^v_{eff}(T_0)}\right)^{1/3}.
\end{align}

Note that the left hand side is a highly non-linear function of $\xi(T)$
since for our model 
\begin{align}
h^{\gamma'}_{\rm eff}(T_h)=\frac{45}{4\pi^4}\int^{\infty}_{x_{h\gamma'}}
\frac{\sqrt{x^2-x_{h\gamma'}^2}}{e^x-1}(4x^2-x_{h\gamma'}^2)dx,\nonumber\\
  h^{D}_{\rm eff}(T_h)=\frac{15}{\pi^4}\int^{\infty}_{x_{hD}}\frac{\sqrt{x^2-x_{hD}^2}}{e^x+1}(4x^2-x_{hD}^2)dx\,,
\end{align}
where $x_{h\gamma'}=m_{\gamma'}/(T_h)= m_{\gamma'}/(\xi(T) T)$
and $x_{hD}= m_{D}/(\xi(T) T)$.\\

In Fig.~\ref{fig:entropy-approx} we give a comparison of  
the analysis of the evolution of the $\xi(T)$ using Eq.(2.15) vs the evolution 
given by the approximation of entropy conservation in co-moving volume
for the visible and the hidden sectors separately. The analysis shows that as $\xi_0$ gets progressively smaller deviations of the 
approximate solutions gets progressively worse and especially in 
the freeze-in region where $\xi_0=0.001$, the deviations of the 
approximate from the exact is huge for temperatures in the visible
sector below $10^5$ GeV. More importantly for any choice of $\xi_0$
in the range $(0,1)$ which includes both the freeze-out and the 
freeze-in regions, the prediction of $\xi_0$ for the approximation
is always inaccurate at the BBN temperature of $\sim 1$ MeV. 
The right panel gives a plot of $\xi$ as a function of the visible sector
temperature for different values of $\delta$ for the case when
$\xi_0=0.001$. Here one finds that the approximation (dashed line)
 gives reasonably accurate result for the case when $\delta=0$,
i.e., there is no kinetic mixing but it gives highly inaccurate results
for the case when $\delta$ in non-vanishing, even as small
as $\delta \sim 10^{-10}$. 

\begin{figure}[h]
    \centering
   \includegraphics[width=0.4\linewidth]{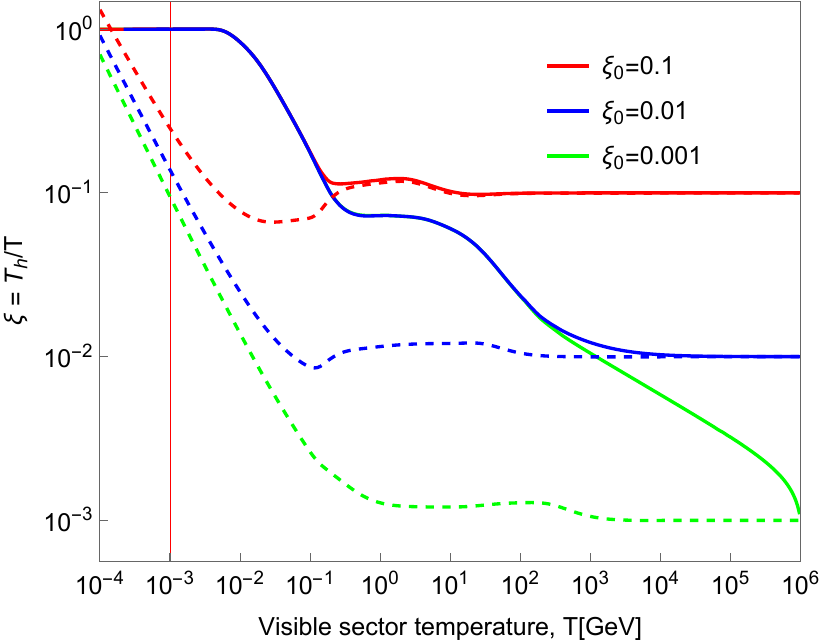}
      \includegraphics[width=0.4\linewidth]{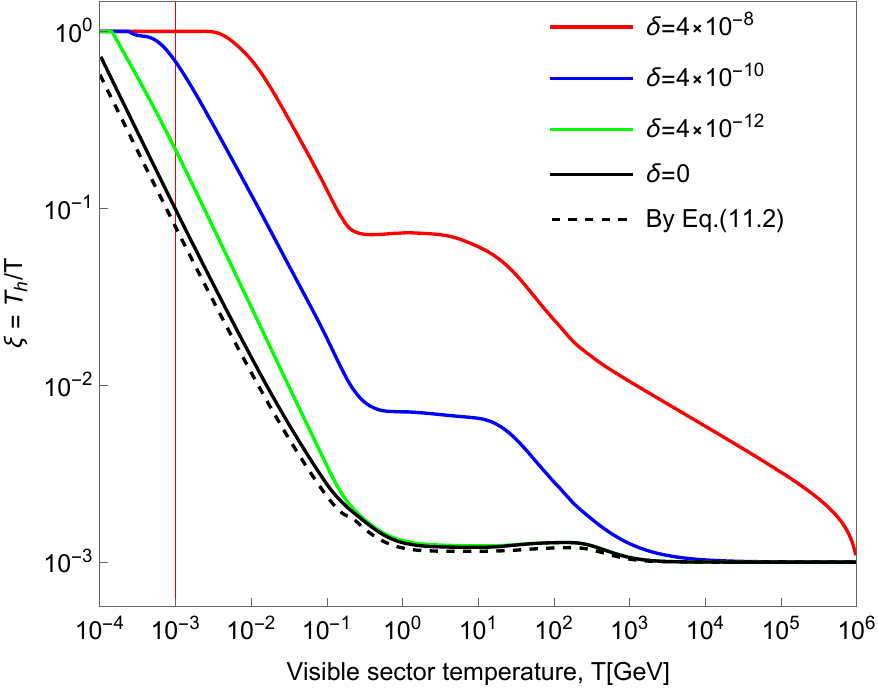}
    \caption{
    Evolution of $\xi(T)$ with different initial condition using Eq.(\ref{DE1}) of this paper (solid) and using the approximation of entropy conservation (dashed). Left panel: Here
$\delta = 4\times 10^{-8}$ and analysis is given for three 
widely different values of $\xi_0$, i.e., $\xi_0=0.001,\xi_0=0.01,
  \xi_0=0.1$.
 Right panel:  Here $\xi_0=0.001$ and an analysis for several different values 
 for $\delta$ in the range $\delta=0$ to $\delta=10^{-8}$ is
 exhibited. The rest of parameters are chosen so that $m_D = 2$ GeV, $m_{\gamma'} = 2$ MeV, $g_X = 0.015$.}
    \label{fig:entropy-approx}
\end{figure}

In the analysis done so far we assumed $M_2 = 0$.
 For generality we consider now the case where we include
 the mass mixing parameter $\epsilon$ along with kinetic mixing $\delta$. Thus we discuss again the thermal 
 evolution when there are both 
kinetic mixing and mass mixing present where we use the relations
given by Eq.(\ref{delta-epsilon-1}) and Eq.(\ref{delta-epsilon-2}) 
and related relations given in \cite{Feldman:2007wj}. In Fig.~\ref{fig:epsilon} we investigate the effect of including $\epsilon$ along $\delta$ on the evolution of $\xi(T)$. 
The left panel is for the case $\epsilon=0. 9\delta$ with 
$\delta=4\times 10^{-8}$ and as expected the evolution for 
different $\xi_0$ shows  a pattern similar to the left panel
of Fig.~\ref{fig:entropy-approx}. 
The right
panel of Fig.~\ref{fig:epsilon} shows that evolutions with different $\epsilon$ follow a similar path at high temperatures but 
begin to separate at $T \sim 10$ GeV.
This separation results in 
 significantly different values of $\xi(T)$ at BBN temperature.
 As expected we find that since $\delta$ and $\epsilon$
 together control the thermal evolution, there is  a significant
 difference in the pattern of evolution here relative to those of  
  Fig.~\ref{fig:entropy-approx}. However, for both
 Fig.~\ref{fig:entropy-approx} and Fig.~\ref{fig:epsilon} 
 one finds that the predictions for $\xi(T)$ given by the 
 approximation equation Eq.(\ref{entropy-con}) shown by dashed curves differs
  by wide margins from the result using 
 Eq.(\ref{DE1}) over wide regions of the parameter space and specifically at BBN temperature. Thus, our conclusion, is that the
entropy conservation approximation separately for the visible and hidden sectors 
in thermal evolution is not suitable for a precision analysis. 

\begin{figure}[h]
    \centering
   \includegraphics[width=0.4\linewidth]{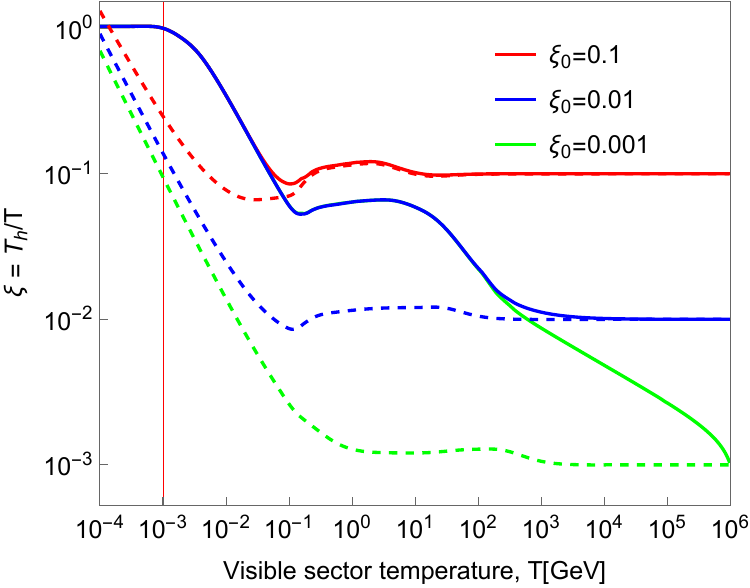}
      \includegraphics[width=0.4\linewidth]{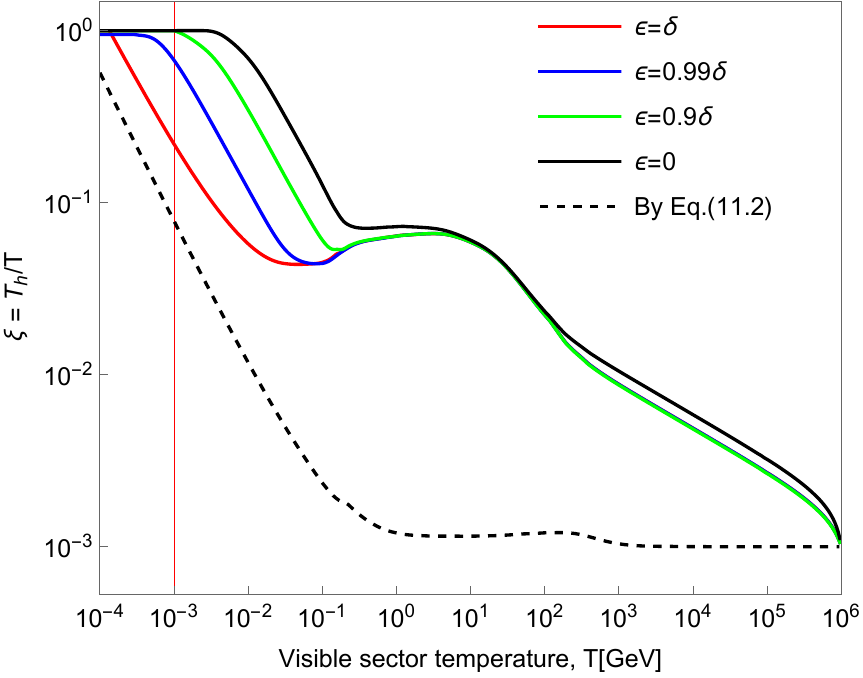}
    \caption{ 
    Evolution of $\xi(T)$ with different initial conditions using Eq.(\ref{DE1}) of this paper (solid) and using the approximation of entropy conservation (dashed). Left panel: Here 
     $\epsilon = 0.9\delta$,
      and the plot is for three different values of $\xi_0$ as shown.
     Right panel: Here $\xi_0=0.001$ and the plots are for
     several different values of $\epsilon$.  For both the left and 
     the right panels the rest of parameters are chosen to be $m_D = 2$ GeV, $m_{\gamma'} = 2$ MeV, $g_X = 0.015$, $\delta = 4\times 10^{-8}$.}
    \label{fig:epsilon}
\end{figure}


 \subsection{On the validity of the total entropy conservation \label{ent2}}
  In the preceding analysis we  discussed the evolution of $\xi(T)$ for the case
  when entropy conservation per comoving volume 
  is assumed separately for the visible and the hidden sectors vs the case when the 
  entropy  conservation is assumed only for their sum. It is found that the deviations between
  the two analysis could be significant,
  as exhibited by Fig. (\ref{fig:entropy-approx}) and Fig.(\ref{fig:epsilon}).
   It is then pertinent to ask the validity of conservation of total entropy since the total
   entropy itself in not conserved either unless various sectors themselves
   equillibrate.  We note that in our analysis the deduction of the evolution equation
   for $\xi(T)$ did not involve any assumptions related to entropy and the only
   place where the conservation of the total entropy was used was in the yield equations.
   For this reason we reconsider the Botzmann equation for the yields without the 
   assumption of total entropy conservation. We focus on the yield equation for the $D$-fermion which constitutes dark matter in the model and the analysis for the 
   yield for the dark photon is very similar. 
   
Thus we start with the Boltzmann equation for the number density $n_D$ which is given by
\begin{align}
    \frac{dn_D}{dt}+3Hn_D = C_D =& \Big[\left<\sigma v\right>_{D\bar{D} \rightarrow i\bar{i} }(T)n_D^{eq}(T)^2 -\left<\sigma v\right>_{D\bar{D} \rightarrow \gamma'\bar{\gamma'} }(T_h)n_D(T_h)^2\nonumber\\
      &+\left<\sigma v\right>_{\gamma'\bar{\gamma'} \rightarrow D\bar{D} }(T_h)n_{\gamma'}(T_h)^2 \Big]
\end{align}
We note now that the equation for the yield $Y_D=n_D/{\bold s}$ without the use of entropy conservation gives 
so that 
\begin{align}
   \frac{dY_D}{dT} 
   &= \frac{1}{\bold s}\frac{dn_D}{dT}-\frac{n_s}{\bold{s}^2}\frac{d\bold s}{dT}\nonumber\\
   & = -\frac{\bold s}{H}\frac{d\rho_v/dT}{4\zeta\rho   - 4\zeta_h\rho_h   + j_h/H}(\frac{C_D}{\bold s^2})
    + \frac{Y_D}{4H\bold s \zeta\rho} (\frac{d\rho}{dT}) \Delta_{\bold s}.\non
    \Delta_{\bold s}&\equiv [\frac{d{\bold s}}{dt}+ 3H{\bold s}]    
    \label{Eq.newYD}
\end{align}
We notice that the set of terms on the right-hand side of Eq.(\ref{Eq.newYD})
involving $C_D$ are exactly what we  have in Eq.(\ref{DE2}). 
Further, the term involving $\Delta_s$ vanishes on using the conservation of 
total entropy and indicates the deviation of the exact equation from the approximate
one where total entropy conservation is assume. A similar analysis holds for the case of the dark photon yield equation. Thus we carry out an analysis using the exact equations
without entropy conservation constraint and compare it with the analysis where entropy conservation is assumed. Results are presented 
in Fig.(\ref{fig:COE}).
\begin{figure}[h]
    \centering
   \includegraphics[width=0.4\textwidth]{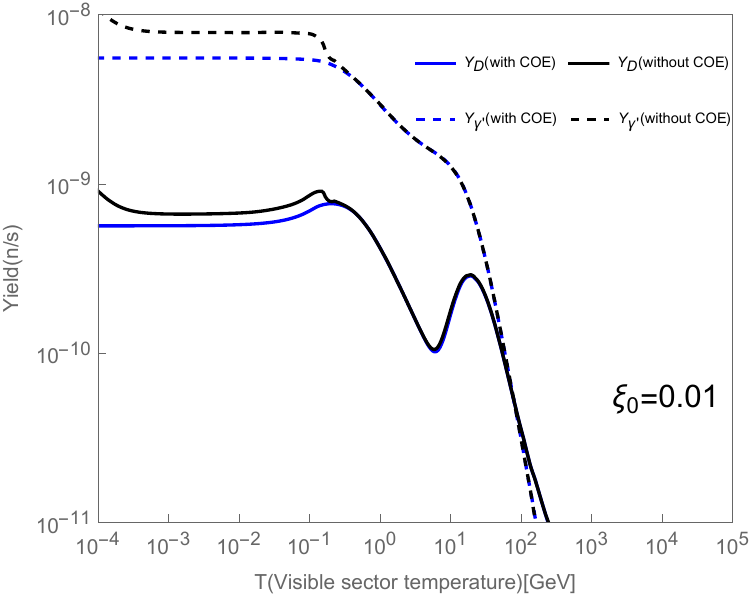}
 \includegraphics[width=0.4\textwidth]{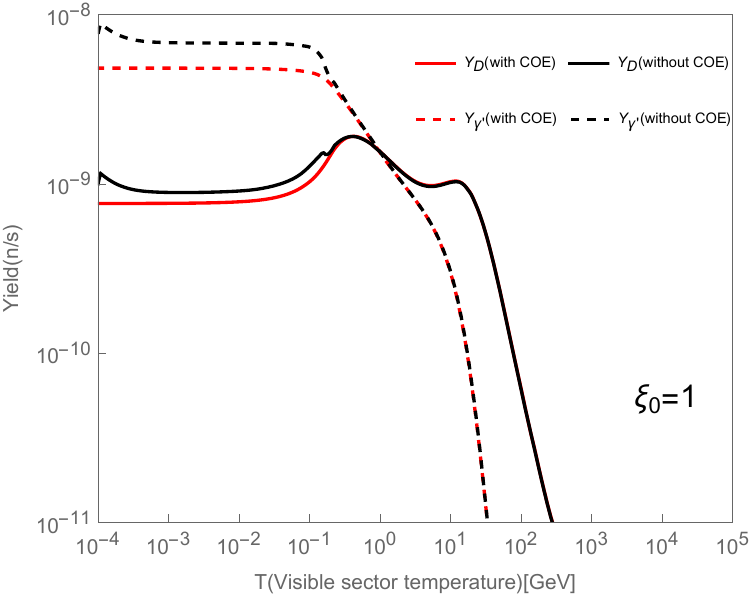}
    \caption{Left panel: Yields of dark fermion and dark photon when $\xi_0 = 0.01$ with and without conversation of entropy (COE). 
    Right panel: Evolution of $\xi(T)$ when $\xi_0 = 1$ with and without COE. The model point we used here is the same as Fig.(\ref{fig:relicchange}), which is model (f) of Table \ref{tab:benchmarks}. }
    \label{fig:COE}
\end{figure}
The analysis of Fig.(\ref{fig:COE}) shows, when  the conservation of entropy (COE) is dropped, the results do not change a lot. Thus the top left panel  for $\xi_0=0.01$ 
shows that the yield $Y_D$ changes by typically within $\sim 15\%$ without inclusion of the entropy conservation constraint.
A similar analysis holds for the case $\xi_0=1$ 
as shown on the right panel of Fig.(\ref{fig:COE}). 
 However, we point out an issue that arises at  very low 
  temperatures. Without COE constraint the yields begin to exhibit an instability
  at low temperature at around $10^{-4}$ GeV. In part this could be due to 
  lack of analytic expressions for the entropy degrees of freedom in the visible 
  sector where on relies on curves or tabulated data 
   (see, e.g.,\cite{Hindmarsh:2005ix,Husdal:2016haj}) because
  of hadronization of quarks and gluons. The instability arises essentially from the
  terms proportional to $\Delta_s$. A proper analysis of this issue is outside the
  scope of this work and a relevant topic for further investigation.


\end{document}